\documentclass[12pt]{amsart}

\usepackage[latin1]{inputenc}
\usepackage{amscd,amsfonts,amsmath,amssymb,amsthm,latexsym}
\usepackage{enumerate,array,srcltx,amsmath}
\usepackage{leqno}
\usepackage[dvips]{graphicx}

\usepackage{caption}
\usepackage{subcaption}
\usepackage{physics}
\usepackage{multirow}

\usepackage{xcolor}

\definecolor{darkblue}{rgb}{0,0,0.545098}
\definecolor{darkgreen}{rgb}{0,0.392157,0}

\usepackage[driverfallback=pdfmx,colorlinks,bookmarksopen,bookmarksnumbered,citecolor=blue, urlcolor=blue]{hyperref}

\theoremstyle{definition}

\theoremstyle{remark}

\numberwithin{equation}{section}

\title[\emph{Castells} modelling long report]{Human towers or \emph{castells} modelling \\ \medskip 158th European Study Groups with Industry\\ (long report)}
\author[\emph{Castells} modelling long report]{J. Antunes$^1$, F. Brosa Planella$^2$, A. D\`oria-Cerezo$^3$, A. March San Jos\'e$^4$,\\ M. Pellicer$^{5}$, A. Rodr\'{\i}guez-Ferran$^{6}$, J. Saludes$^{7}$}
\thanks{$^1$ Instituto Superior T\'ecnico, Lisboa. E-mail address: jantunes@ctn.tecnico.ulisboa.pt}
\thanks{$^2$ University of Warwick. E-mail address: Ferran.Brosa-Planella@warwick.ac.uk}
\thanks{$^3$ Universitat Polit\`ecnica de Catalunya. E-mail address: arnau.doria@upc.edu}
\thanks{$^4$ Universitat Aut\`onoma de Barcelona. E-mail address: aleixmarchsj@gmail.com}
\thanks{$^{5}$ Universitat de Girona. E-mail address: marta.pellicer@udg.edu}
\thanks{$^{6}$ Universitat Polit\`ecnica de Catalunya. E-mail address: antonio.rodriguez-ferran@upc.edu}
\thanks{$^{7}$ Universitat Polit\`ecnica de Catalunya. E-mail address: jordi.saludes@upc.edu}
\makeatother

\date{\today}

\newcommand{\castell}[1][]{\emph{castell#1}}

\begin{document}
\small{\maketitle}
\begin{abstract}
Human towers or \emph{castells} are human structures played in festivals mainly in Catalonia. These unique cultural and traditional displays have become very popular in the last years, but they date from the XVIII century. On 2010 they became part of the Unesco Representative List of the Intangible Cultural Heritage of Humanity. 

Safety is very important in the performance of \emph{castells}. To this end, it is crucial to understand the mechanisms that allow a \emph{castell} to be built and, more importantly, the factors that may cause its collapse. This work is focused on the mechanical aspects that make a \emph{pilar} (the simplest structure in the \emph{castells}) to stand. We suggest three different but complementary approaches for the running stage of a \emph{pilar} (stage where it has been built and has not yet collapsed): the $N$-link pendulum as a first dynamical model, the response of the \emph{castellers} as a control problem, and a static analysis to capture the feasibility of a given configuration. We include some preliminary simulations to better understand the previous approaches, which seem to match with qualitative perceptions that \emph{castellers} have of a \emph{castell}. Possible future developments are also discussed. To our knowledge, this work represents the first one to study the \emph{castells} from a mechanical point of view.
\end{abstract}

\section{Introduction and statement of the problem}\label{sec:intro}

\begin{figure}[htpb]
\begin{center}
\includegraphics[width=6cm]{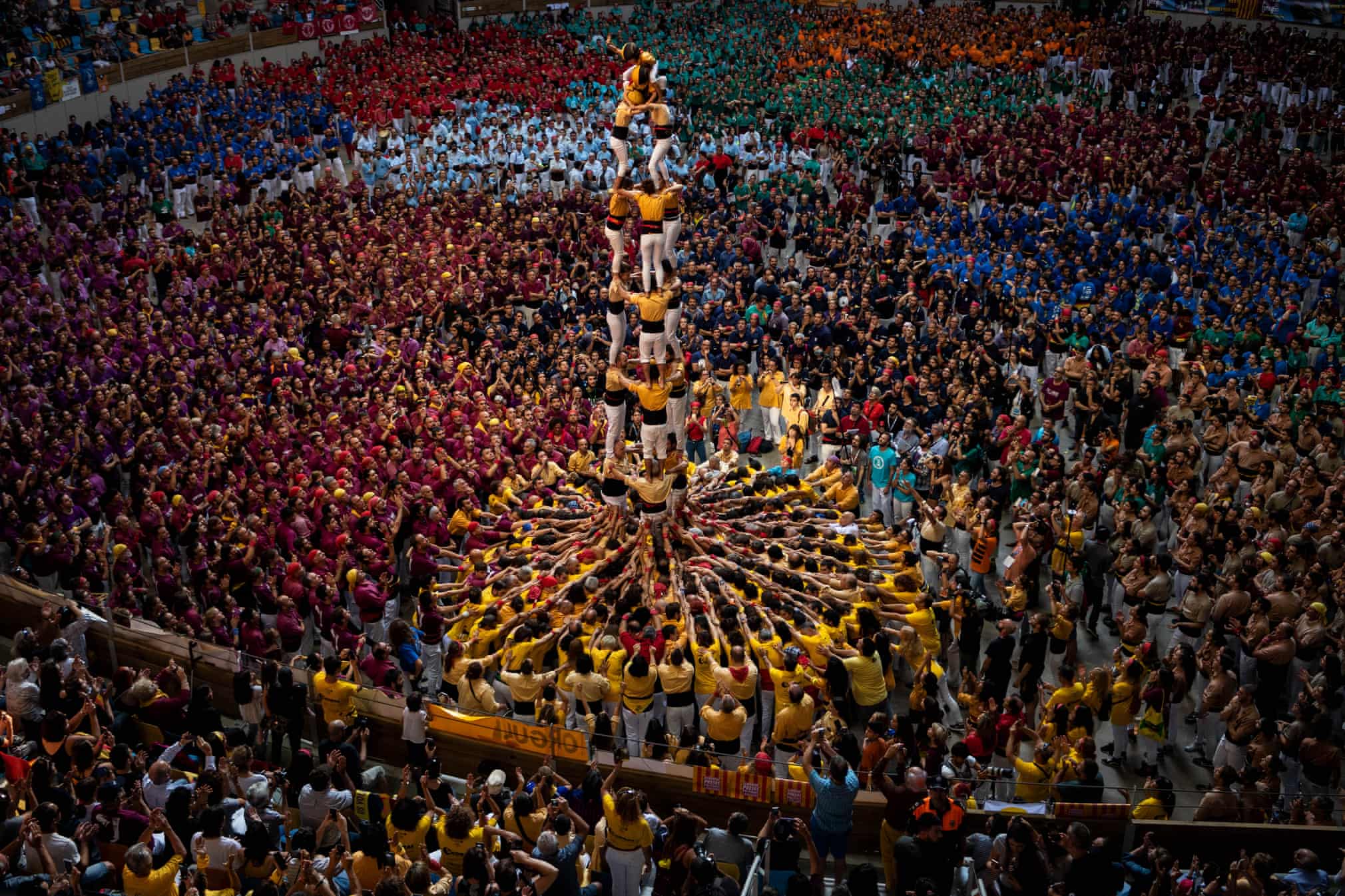} \hspace{1cm} \includegraphics[width=5.7cm]{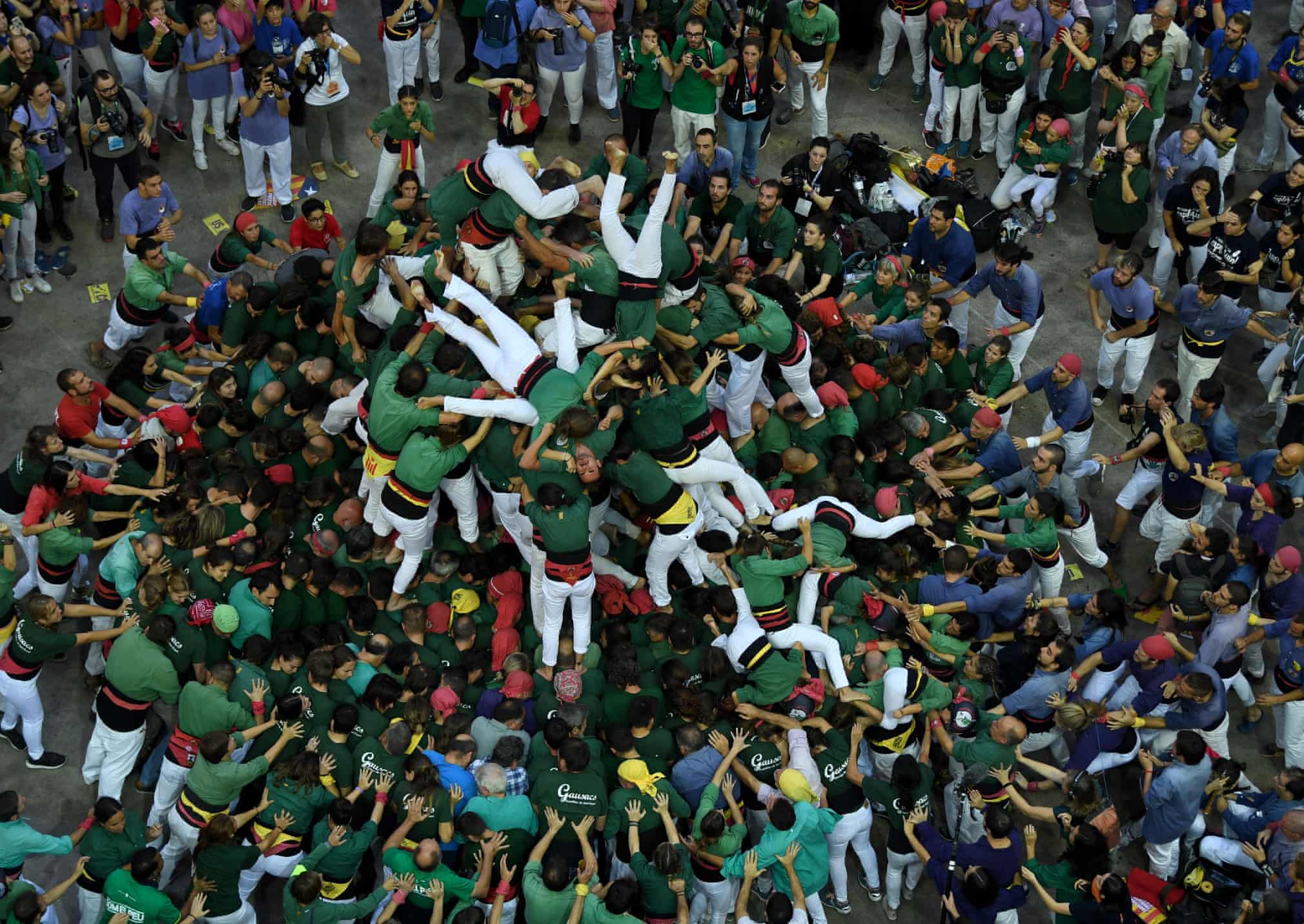}
\caption{\emph{Tres de vuit} (3d8) of the \emph{colla} Bordegassos de Vilanova (left), image over the \emph{pinya} once a \emph{castell} has fallen down (right). Photos at the biannual \emph{castells} competition at Tarragona, from \href{https://www.theguardian.com/world/gallery/2018/oct/08/castellers-human-towers-of-catalonia}{The Guardian}.}\label{fig:castells}
\end{center}
\end{figure}

\emph{Castells}\footnote{The emphasized terms are technical terms and, hence, we have kept their form in Catalan. A Glossary with all of them is included in Appendix \ref{sec:appendix}} are human towers (see \cite{videoUNESCO,CC,Wiki} and Figure \ref{fig:castells}) built traditionally at festivals in Catalonia,  the Balearic Islands and the Valencian Community.
It started from the \emph{Ball de Valencians} in Valls (first documented in 1712) and spread over the course of the 18th century to other towns in the area. Increasing its popularity over the last years, recently there are accounted around a hundred teams or \emph{colles} around the world, most of them in Catalonia (see \cite{CC}). 

In a \emph{castell} a number of people (\emph{castellers}) perform a human structure climbing one on the top of each other. A \emph{castell} succeeds when a child called \emph{enxaneta} has climbed to the top of the \emph{castell}, crossed the top of the structure (raising their hand at this moment) and the \emph{castell} has been successfully disassembled (or \emph{descarregat}) afterwards. If the \emph{castell} falls down once the \emph{enxaneta} has arrived at the top of it, it is said to be only \emph{carregat}. The basic structure of a \emph{castell }consists of three parts: the \emph{pinya} (or basis), the \emph{tronc} (the vertical part of the \emph{castell}) and the \emph{pom de dalt} (the top three levels of the \emph{castell}, with the \emph{enxaneta} at the top one).

Most \emph{castells} are succesful, but some of them end up collapsing. The Coordinadora de Colles Castelleres de Catalunya (CCCC), the \emph{colles} coordinating body, has as one of its tasks to help the \emph{colles} to improve their safety while doing \emph{castells}. The current work lays in this area. It is part of the largest project suggested by the CCCC consisting of determining the area of the \emph{pinya} where is more likely that \emph{castellers} impact when a \emph{castell} collapses. As we said, this is important as it would allow the CCCC to increase the safety of the \emph{pinya} members: it would allow to quantify the minimum required size of a \emph{pinya} for preventing people falling directly on the floor, and also to determine the most \textsl{dangerous} areas to prescribe, if necessary, additional external protection devices (such as spine protection elements, currently under study).

\begin{figure}[h]
\centering
\scalebox{0.70}{
\def\svgwidth{12.0cm}
\begingroup%
    \setlength{\unitlength}{\svgwidth}%
      \begin{picture}(1,0.65)%
    \put(0,0){\includegraphics[width=\unitlength]{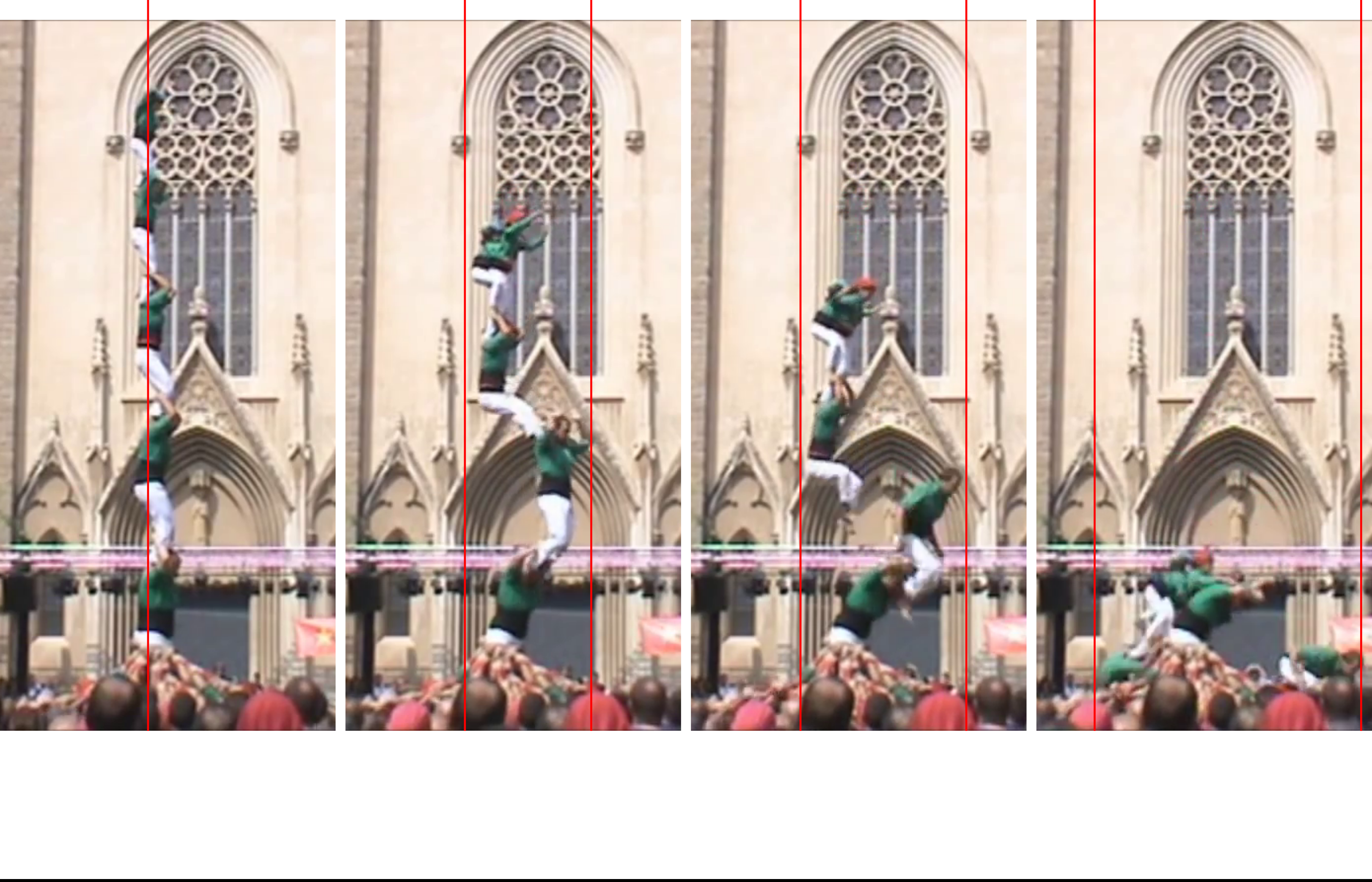}}%
    	\put(0.12,0.05){\color[rgb]{0,0,0}\makebox(0,0)[cb]{\smash{\begin{tabular}{c}Stable\\stage\end{tabular}}}}%
	\put(0.37,0.05){\color[rgb]{0,0,0}\makebox(0,0)[cb]{\smash{\begin{tabular}{c}Failure\\stage\end{tabular}}}}%
	\put(0.62,0.05){\color[rgb]{0,0,0}\makebox(0,0)[cb]{\smash{\begin{tabular}{c}Falling\\stage\end{tabular}}}}%
    	\put(0.87,0.05){\color[rgb]{0,0,0}\makebox(0,0)[cb]{\smash{\begin{tabular}{c}Impact\\stage\end{tabular}}}}%
    \end{picture}%
  \endgroup%
}
\caption{Stages when a \emph{castell} collapses and their associated risky areas (in red). The photo corresponds to a \emph{pilar de 6 carregat (pd6c)}, from  Castellers de Sabadell, 2012.}\label{FigureStages}
\end{figure}
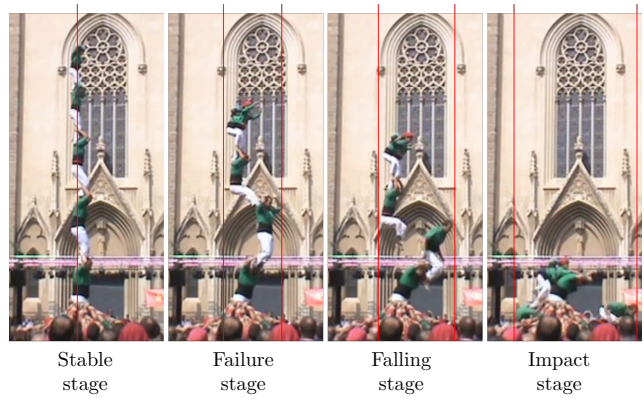

To this end, our approach split the above problem in the following stages (see Figure \ref{FigureStages}):
\begin{itemize}
\item The \textbf{standing or running stage}, containing both the \textbf{stable} and \textbf{failure} stages in the building of a \emph{castell}. It hence includes the admissible dynamics of a \emph{castell} when it is being built and the analysis of the limits where \emph{castellers} are able to remain together and the mechanisms that allow it or the factors that make a \emph{castell} to collapse.
\item The \textbf{falling stage} starts when lateral forces vanish and \emph{castellers} are only affected by gravity. Then, gravity and both plastic and elastic impacts between \emph{castellers} need to be considered.
\item Finally, the \textbf{impact stage} contain the first impact to the \emph{pinya} and all the subsequent impacts until all \emph{castellers} stop on it.
\end{itemize}

This approach allows to define a risky area related to each stage as well as to understand the mechanisms that are present in each, and provides the initial conditions for the next stage. 

The present work is focused on the study of the running stage and the operating capability of a \emph{castell} within it. The importance of this work is twofold: apart from understanding the dynamics that allow a \emph{castell} to be build, we will try to capture and understand the main mechanisms and factors that make a \emph{castell} to be successful or to collapse. The next stages would analyse how and where the \emph{castellers} of the \emph{tronc} fall down, once the structure of the \emph{castell} is broken. Although Biomechanics is a wide field of study (with plenty of references in the literature) we have to say that, to our knowledge, this is the first work on the modelling of \emph{castells} from a mechanical point of view. In this sense, the few accurate scientific studies in the literature are provided in \cite{Roset, Roset2}.

As we explain just below, in the present work the analysis is focused on the \emph{pilar} structure, composed by just one \emph{casteller} at each level of the \emph{tronc}. Of course, in future works this analysis should be extended to more complex 3D structures.

\vspace{0.2cm}
\noindent\textbf{The pillar: a paradigm of most modelling issues.} \emph{Castells} can display many configurations, ranging from simple \emph{pilars}, see Figure \ref{fig:diagram_pilar}(A), to very complex 3D structures (see Figure \ref{fig:castells}). After carefully scrutinizing the addressed problem, we decided to focus this preliminary work on a generic \emph{pilar}, with $N$ levels. Actually, the geometry of such structures is the simplest one among the \emph{castellers} repertoire (we can think them as 1D structures) and, moreover, they typically display near-planar motions (hence, 2D movement). In spite of this simplicity, the essential features of all \emph{castells} are present when modelling \emph{pilars}, namely the need for (a) both static and dynamical structural models, (b) neuro-muscular control laws, (c) balancing criteria, (d) \emph{castell} disruption criteria, and (e) a realistic post-failure dynamical formulation. Only after all these aspects have been mastered and calibrated on the simplest \emph{pilar} structures, can one attempt to apply the developed computational strategies to more complex \emph{castells} displaying 3D motions.

This work first starts with the dynamical approach (Sections \ref{sec:dynam} and \ref{sec:3-link_whole}) and then focuses on a static balance approach (Section \ref{sec:static}). This may seem contrary to the construction of the models, but we will see how this makes sense and, moreover, how it allows us to start with a 1-link \emph{casteller} model and continue with a more realistic 3-link one.

In Section~\ref{sec:dynam} we present a first dynamical model where we think the  pillar as an inverted pendulum with several links or bars (each of this link representing a \emph{casteller}) stabilized by restoring torques that are applied to their joints. 
First, for the uncontrolled dynamics, the Lagrangian formulation for the problem is proposed based on the link angles and angular velocities. Although this formulation may not be new, it has to be emphasized that the point of view adopted in this section allows to write the corresponding dynamical system in a very compact and straightforward way, which is important as the number of equations may be large.

In Section \ref{sec:3-link_whole} we look for a more realistic model of a \emph{casteller}, which can capture the bent-knee position often used when performing human towers. Therefore, we move from one to three links per \emph{casteller}. This new model raises a few questions that need to be considered. The first one are the biomechanical restrictions in the internal joints, as they cannot be bent beyond a certain angle. We also consider the restoring torques that allow the \emph{casteller} to stand in its desired position as a control problem, where the reaction times of the \emph{casteller} may be included as time delays. This leads us to consider the balance criteria imposed by a \emph{casteller} to maintain their standing position. Finally, we study the dynamics of this bent-knee position: we run simulations for the behaviour under an initial perturbation of different positions. Some preliminary simulations seem to validate the benefits of the bent-knee position, as it allows a faster return to the desired positions, giving as well a better reaction capability to the \emph{casteller}.

In Section~\ref{sec:static} we take another approach. While in the previous sections we have considered a dynamical point of view (first for a 1-link \emph{casteller} and then for a 3-link one) we now consider a static and fixed configuration for a \emph{pilar} and compute the conditions for the restoring torques to ensure its static balance. 
These computations suggest some quality indices showing how far is a given configuration from collapsing (and which will be the effort to avoid it). These indices are the mass center eccentricity (from the perfectly vertical position), and what we call the folding index and the suffering index, which measure how \textsl{bent} are the joints of the \emph{castellers} and how much effort their joint muscles are applying, respectively. As examples of them, we numerically evaluate these indicators on configurations taken from two videos of pillars (one successful, the other not).

As future considerations (among others) it is clear that accurate biomechanical data is needed to validate the approaches above and go further, as much as good positional data on real \castell[s]. The control problem that follows from an active \castell[er] is another question that with no doubt is present in reality and, although it is considered in the present work, it should definitely be developed in a future line of research in this project. Once the current approaches are fully developed for the \emph{pilar} structure, the next step is of course to extend and validate the model to more complex constructions. These and other conclusions, and future lines of research are considered in Section \ref{sec:future}.

Finally, we include a glossary of the most common terms in the world of \emph{castells}, for a better comprehension of the terminology (see Appendix \ref{sec:appendix}).

\section{A basic dynamical point of view: the $N$-link inverted pendulum}\label{sec:dynam}

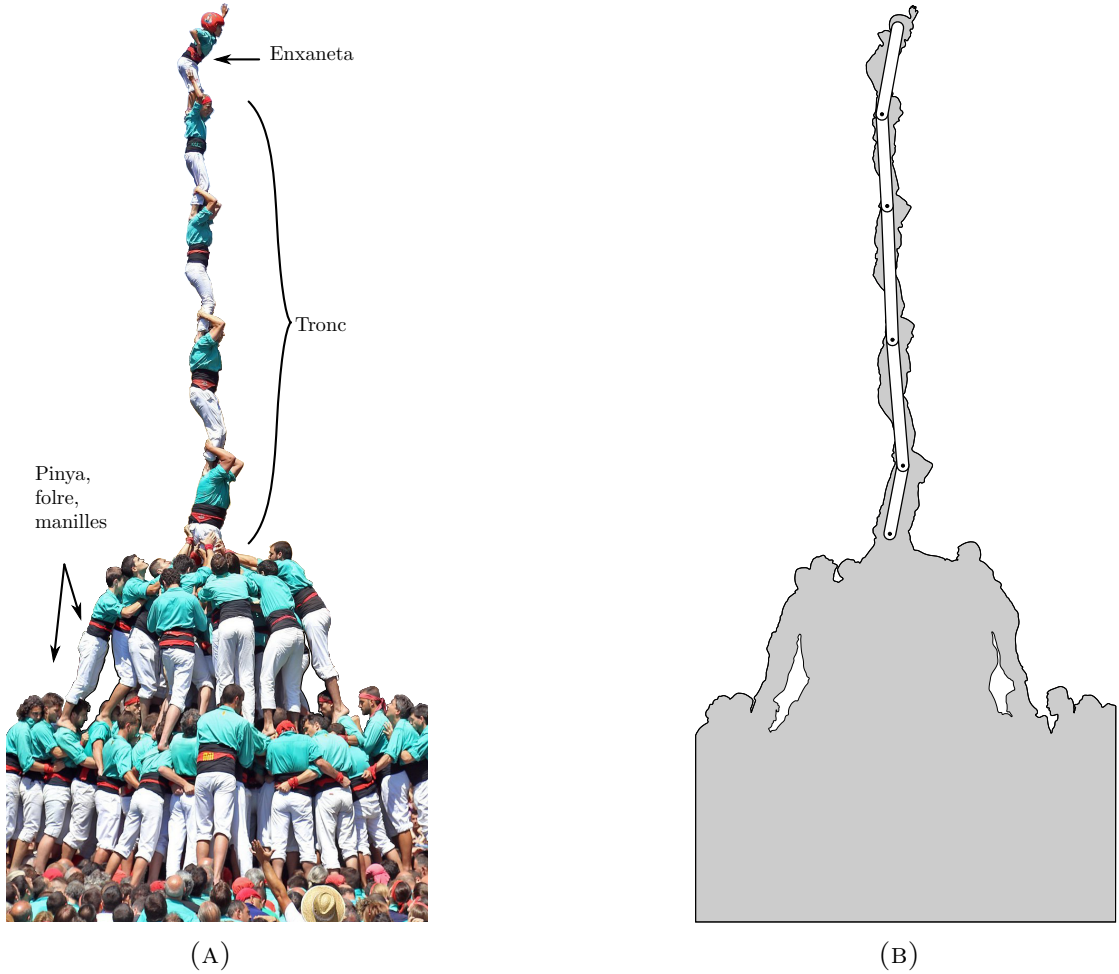
\begin{figure}
\centering
\begin{subfigure}[b]{0.45\textwidth}
	\centering	
	\scalebox{0.70}{
    \def\svgwidth{8.0cm}
    \begingroup%
    \setlength{\unitlength}{\svgwidth}%
    \begin{picture}(1,2.18385265)%
        \put(0,0){\includegraphics[width=\unitlength]{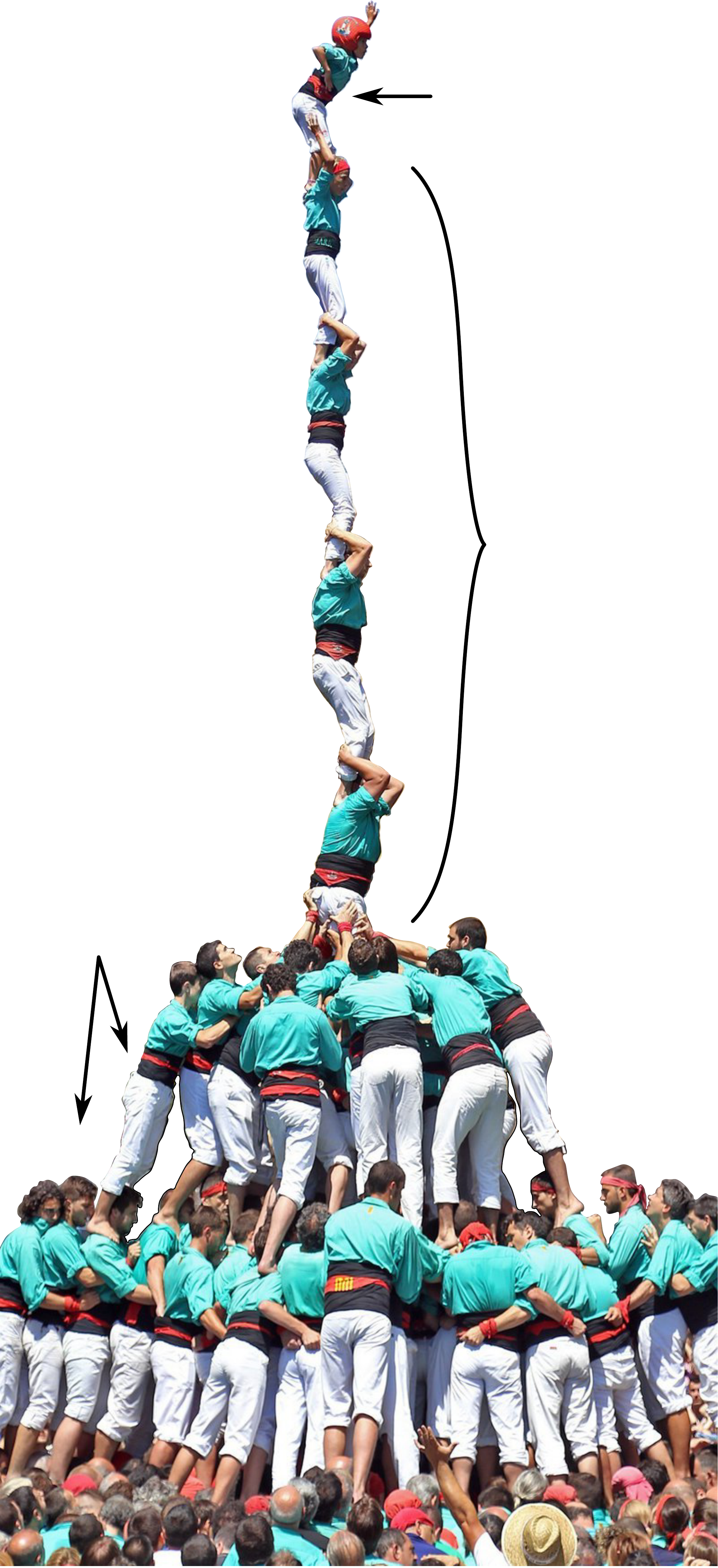}}%
        \put(0.625,2.05){\color[rgb]{0,0,0}\makebox(0,0)[lb]{\smash{Enxaneta}}}%
        \put(0.04592285,0.99385986){\color[rgb]{0,0,0}\makebox(0,0)[lb]{\smash{\begin{tabular}{l} Pinya,\\ folre,\\ manilles\end{tabular}}}}%
        \put(0.68529053,1.40712891){\color[rgb]{0,0,0}\makebox(0,0)[lb]{\smash{Tronc}}}%
    \end{picture}%
    \endgroup%
    }
    \caption{}
\end{subfigure}\hfill
\begin{subfigure}[b]{0.45\textwidth}
	\centering	
	\scalebox{0.70}{
    \def\svgwidth{8.0cm}
    \begingroup%
    \setlength{\unitlength}{\svgwidth}%
    \begin{picture}(1,2.18385265)%
        \put(0,0){\includegraphics[width=\unitlength]{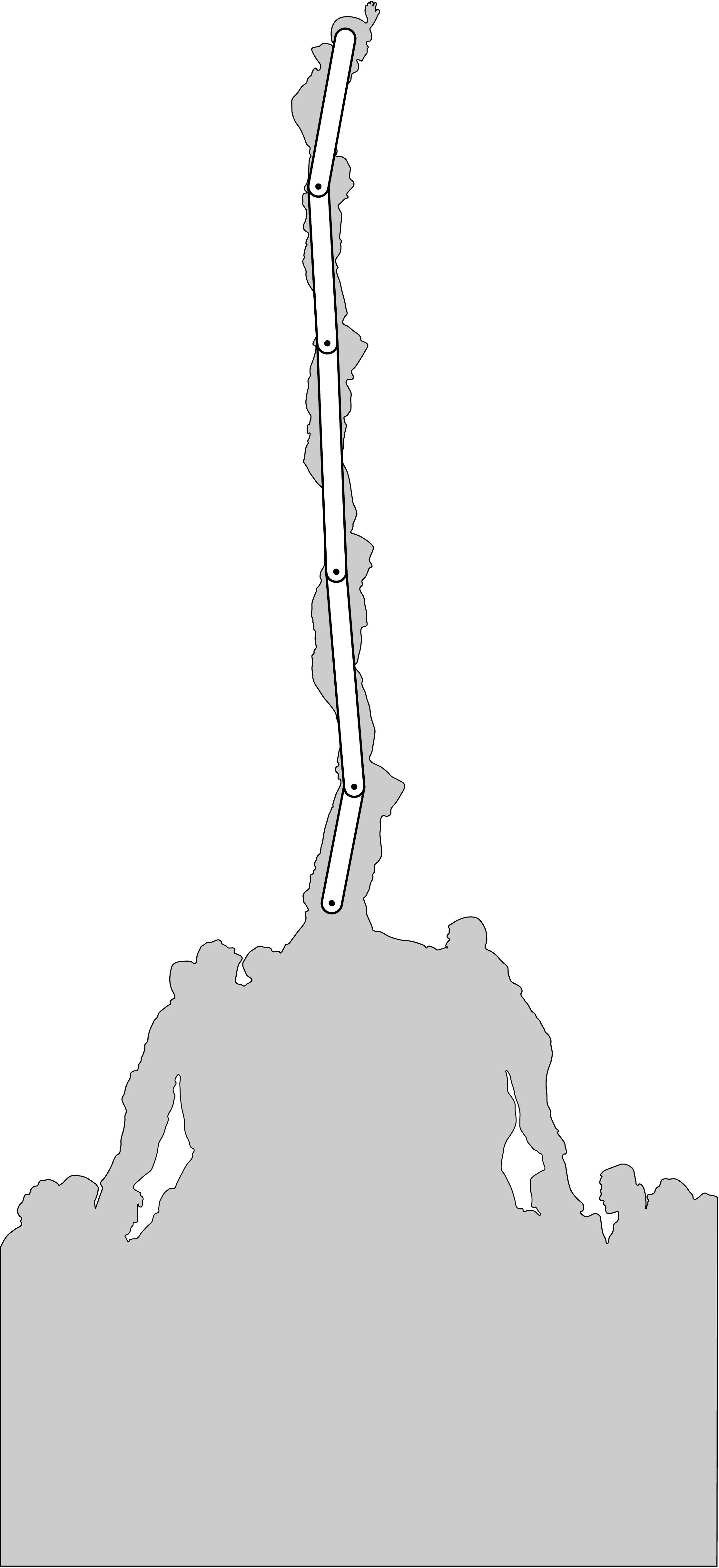}}%
    \end{picture}%
    \endgroup%
    }
	\caption{}
\end{subfigure}
\caption{Example of a real pilar (A) and the diagram with the corresponding approach as an $N$-link inverted pendulum (B). \emph{Pilar de 8 amb folre i manilles (pd8fm)}, Castellers de Vilafranca, Matar\'o, 2014 (Photo: Juanma Ramos,  \href{https://www.elpuntavui.cat/societat/article/5-societat/762686-santes-extraordinaries.html}{El Punt Avui}).}
\label{fig:diagram_pilar}
\end{figure}

In this section we give a non-linear formulation of the assembled $N$-link pillar.
As illustrated in Figure \ref{fig:diagram_pilar}, the \emph{castell} configuration addressed here is the simplest possible, and mimics a \emph{pilar} structure with an arbitrary number $N$ of ``stacked'' \emph{castellers}. Each element $i = 1,2, \dots ,N$ is modelled through a rigid bar with total length $l_i$ and distributed mass, with an equivalent total mass $m_i$ ``assigned'' to the center of mass at local axial coordinate $G_i = c_i l_i$. If one assumes that the bars masses are uniformly distributed, then the relative local axial coordinate of the center of mass of each element is given by $c_i = G_i/l_i$, while the moment of inertia about the center of mass is $J_{i} = m_i l_i^2/12$.

\begin{figure}
\centering	
	\scalebox{0.75}{
    \def\svgwidth{4.0cm}
    \begingroup%
    \setlength{\unitlength}{\svgwidth}%
    \begin{picture}(1,1.88244351)%
        \put(0,0){\includegraphics[width=\unitlength]{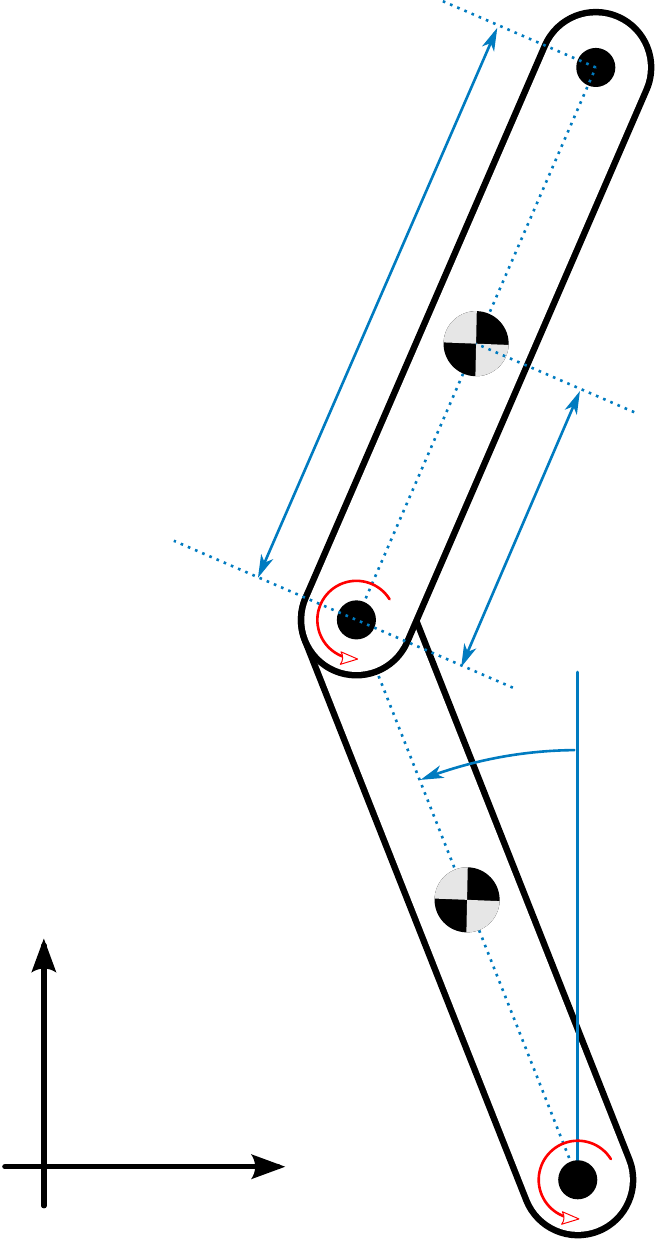}}%
        \put(0.35,0.02){\color[rgb]{0,0,0}\makebox(0,0)[lb]{$x$}}%
        \put(-0.02,0.4){\color[rgb]{0,0,0}\makebox(0,0)[lb]{$y$}}%
        \put(0.75,0.78){\color[rgb]{0,0.498,0.68}\makebox(0,0)[lb]{$\theta_i$}}
        \put(0.65,0.05){\color[rgb]{1,0,0}\makebox(0,0)[lb]{$\Gamma_i$}}
        \put(0.35,0.46){\color[rgb]{0,0,0}\makebox(0,0)[lb]{$m_i,J_i$}}
        \put(0.40,1.50){\color[rgb]{0,0.498,0.68}\makebox(0,0)[lb]{$l_{i+1}$}}
        \put(0.25,0.91){\color[rgb]{1,0,0}\makebox(0,0)[lb]{$\Gamma_{i+1}$}}
        \put(0.85,1.05){\color[rgb]{0,0.498,0.68}\makebox(0,0)[lb]{$c_{i+1}l_{i+1}$}}
    \end{picture}%
    \endgroup%
    }
\caption{Scheme with the proposed notation. We recall that $\Gamma_i$, $\theta_i$, $J_i$ stand for the restoring moment, the rotation angle and the moment of inertia of bar $i$, respectively. We also recall that $c_i,l_i$ stand for the relative location of the center of mass and the total length of the bar $i$, respectively.}
\label{fig:2barsscheme}
\end{figure}

For the fully assembled structure, the bars are articulated at their extremities, each bar being described in terms of the corresponding rotation angle $\theta _i(t)$ (measured counterclockwise from the vertical). Because any inverted pendulum is inherently unstable, in order to stabilize this $N$-link pendulum, it will be assumed that equivalent restoring moments $\Gamma_i(t)$ are applied by the \emph{castellers} legs and arms at the successive articulations. See the used notation in Figure \ref{fig:2barsscheme}.

Referring to Figure \ref{fig:diagram_pilar}(B), the fully assembled $N$-link \emph{pilar} may be formulated in terms of the generalized coordinates $\theta_i(t)$, with $i = 1,2, \dots ,N$. The nonlinear motion equations of the inverted multiple pendulum will be developed using the Lagrange formulation, with respect to the generalized displacements $\theta_i(t)$ and velocities $\dot{\theta}_i(t)$ (see footnote\footnote{For the sake of simplicity, from now on and when clear from the context, the time argument $(t)$ in the variables is omitted.}):
\begin{equation}\label{eq:lagrangian}
    \dv{}{t}\left( \pdv{T}{\dot{\theta}_i} \right) - \pdv{T}{\theta_i} + \pdv{V}{\theta_i} = \Gamma_i, \quad i = 1,2,\dots,N.
\end{equation}

The potential energy $V$ and the kinetic energy $T$ are given by
\begin{subequations}\label{eq:pot_kin_energies}
\begin{align}
    V(t) &= g \sum_{i=1}^N m_i y_i,\\
    T(t) &= \frac{1}{2} \sum_{i=1}^N \left( m_i |\vb*{v}_i|^2  + J_{i} \dot{\theta}_i^2 \right),
\end{align}
\end{subequations}
where the locations of the centers of mass of the links are given by
\begin{subequations}
\begin{align}
    x_i(t) &= -c_i l_i \sin \theta_i - \sum_{j=1}^{i-1} l_j \sin \theta_j,\\
    y_i(t) &= c_i l_i \cos \theta_i + \sum_{j=1}^{i-1} l_j \cos \theta_j,
\end{align}
\end{subequations}
and the velocities by 
\begin{equation}\label{eq:velocity}
    \vb*{v}_i(t) = \begin{pmatrix}\dot{x}_i\\\dot{y}_i\end{pmatrix}
    =-\begin{pmatrix}c_i l_i \dot{\theta}_i \cos \theta_n + \sum_{j=1}^{i-1} l_j \dot{\theta}_j \cos \theta_j\\c_i l_i \dot{\theta}_i \sin \theta_i + \sum_{j=1}^{i-1} l_j \dot{\theta}_j \sin \theta_j\end{pmatrix}.
\end{equation}

Then, we obtain from \eqref{eq:pot_kin_energies}-\eqref{eq:velocity} 
\begin{subequations}\label{eq:pot_kin_energies2}
\begin{equation}
    V(t) = g \sum_{i=1}^N m_i \left( c_i l_i \cos \theta_i + \sum_{j=1}^{i-1} l_j \cos \theta_j \right),\\
\end{equation}
\begin{equation}
    \begin{aligned}
    T(t) &= \frac{1}{2} \sum_{i=1}^N m_i \left(c_i l_i \dot{\theta}_i \cos \theta_i + \sum_{j=1}^{i-1} l_j \dot{\theta}_j \cos \theta_j \right)^2 \\
    & \quad + \frac{1}{2} \sum_{i=1}^N m_i \left( c_i l_i \dot{\theta}_i \sin \theta_i + \sum_{j=1}^{i-1} l_j \dot{\theta}_j \sin \theta_j \right)^2 + \frac{1}{2} \sum_{i=1}^N J_{i} \dot{\theta}_i^2,
    \end{aligned}
\end{equation}
\end{subequations}
and, replacing \eqref{eq:pot_kin_energies2} into \eqref{eq:lagrangian} we can obtain the motion equations.

A convenient systematic manner to obtain the motion equations of the assembled system is provided in \cite{Kajita}. First, in order to state the kinematic variables, the coordinates of the centers of mass of the links are expressed in terms of the angles $\theta_i(t)$ at the joints. Then, we can write the positions as
\begin{subequations}
\begin{align*}
    \vb*{x}(t) &= -\vb*{\Psi} \sin\vb*{\theta},\\
    \vb*{y}(t) &= \vb*{\Psi} \cos\vb*{\theta},
\end{align*}
\end{subequations}
where $\vb*{x}(t)=(x_1,\ldots,x_N)^T$, $\vb*{y}(t)=(y_1,\ldots,y_N)^T$, $\vb*{\theta}(t)=(\theta_1,\ldots,\theta_N)^T$,
and with $\sin$ and $\cos$ applied elementwise to $\vb*{\theta}$. The transformation matrix $\vb*{\Psi}$ is defined as
\begin{equation*}
    \vb*{\Psi} =
    \begin{pmatrix}
    c_1 l_1 & 0 & 0 & \cdots & 0 & 0 \\
    l_1 & c_2 l_2 & 0 & \cdots & 0 & 0 \\
    l_1 & l_2 & c_3 l_3 & \cdots & 0 & 0 \\
    \vdots & \vdots & \vdots & \ddots & \vdots & \vdots \\
    l_1 & l_2 & l_3 & \cdots & c_{N-1} l_{N-1} & 0 \\
    l_1 & l_2 & l_3 & \cdots & l_{N-1} & c_N l_N
    \end{pmatrix}.
\end{equation*}

Then, the velocities of the links' centers of mass are computed from the time-derivatives
\begin{subequations}
\begin{align*}
    \dot{\vb*{x}} &= -\vb*{\Psi} \dv{}{t} \left(\sin  \vb*{\theta} \right) = - \vb*{\Psi} \vb{D}_\mathrm{c} \vb*{\dot{\theta}},\\
    \dot{\vb*{y}} &= \vb*{\Psi} \dv{}{t} \left(\cos \vb*{\theta} \right) = - \vb*{\Psi} \vb{D}_\mathrm{s} \vb*{\dot{\theta}},
\end{align*}
\end{subequations}
with the $N\times N$ diagonal matrices $\vb{D}_\mathrm{c}(t)=\text{diag}(\cos\vb*{\theta})$ and $\vb{D}_\mathrm{s}(t)=\text{diag}(\sin\vb*{\theta})$.

Now, from the previous results, we can compute the potential and kinetic energies in matrix/vector form
\begin{subequations}
\begin{align*}
    V(t) &= g \vb*{m}^T \vb*{y},\\
    T(t) &= \frac{1}{2} \left( \vb*{\dot{x}}^T \vb{D}_m \vb*{\dot{x}} + \vb*{\dot{y}}^T \vb{D}_m \vb*{\dot{y}} + \vb*{\dot{\theta}}^T \vb{D}_J \vb*{\dot{\theta}} \right),
\end{align*}
\end{subequations}
hence
\begin{subequations}\label{eq:pot_kin_energies3}
\begin{align}
    V(t) &= g \vb*{m}^T \vb{\Psi} \cos \vb*{\theta},\\
    T(t) &= \frac{1}{2} \vb*{\dot{\theta}}^T \left( \vb{D}_\mathrm{c} \vb{\Psi}^T \vb{D}_m \vb{\Psi} \vb{D}_\mathrm{c} + \vb{D}_\mathrm{s} \vb{\Psi}^T \vb{D}_m \vb{\Psi} \vb{D}_\mathrm{s} + \vb{D}_J \right) \vb*{\dot{\theta}},
\end{align}
\end{subequations}
with $\vb*{m}=(m_1,\ldots,m_N)^T$, $\vb{D}_m=\text{diag}(\vb*{m})$ and $\vb{D}_J=\text{diag}(J_1,\ldots,J_N)$.

Then, the dynamical equations are obtained replacing \eqref{eq:pot_kin_energies3} in to the Lagrange equation in vector form
\begin{equation*}
    \dv{}{t} \left( \pdv{T}{\vb*{\dot{\theta}}} \right) - \pdv{T}{\vb*{\theta}} + \pdv{V}{\vb*{\theta}} = \vb*{\Gamma},
\end{equation*}
with the vector of restoring torques defined as $\vb*{\Gamma}(t)=(\Gamma_1,\ldots,\Gamma_N)^T$.

For details of the matrix/vector derivative computations see \cite{Seber}. The following nonlinear dynamical formulation is obtained for the assembled pillar
\begin{equation}\label{eq:n-link_pilar_D}
    \vb{M}\left(\vb*{\theta}\right)\vb*{\ddot{\theta}}+ \vb{D}(\vb*{\theta}) \vb*{\dot{\theta}^2}+
    \vb*{g}\left(\vb*{\theta}\right) = \vb*{\Gamma}
\end{equation}
hence, using a more convenient notation
\begin{equation}\label{eq:n-link_pilar}
    \vb{M}\left(\vb*{\theta}\right)\vb*{\ddot{\theta}}+ \vb{C}(\vb*{\dot{\theta}},\vb*{\theta}) \vb*{\dot{\theta}}+
    \vb*{g}\left(\vb*{\theta}\right) = \vb*{\Gamma}
\end{equation}
where
\begin{align*}
    \vb{M}\left(\vb*{\theta}\right)&= \vb{D}_\mathrm{c} \vb{\Psi}^T \vb{D}_m \vb{\Psi} \vb{D}_\mathrm{c} + \vb{D}_\mathrm{s} \vb{\Psi}^T \vb{D}_m \vb{\Psi} \vb{D}_\mathrm{s} + \vb{D}_J \\
    \vb{C}(\vb*{\dot{\theta}},\vb*{\theta})&=\vb{D}(\vb*{\theta})\text{diag}(\vb*{\dot{\theta}}) = \left[ \vb{D}_\mathrm{s} \vb{\Psi}^T \vb{D}_m \vb{\Psi} \vb{D}_\mathrm{c} - \vb{D}_\mathrm{c} \vb{\Psi}^T \vb{D}_m \vb{\Psi} \vb{D}_\mathrm{s} \right] \text{diag}(\vb*{\dot{\theta}}) \\
    \vb*{g}\left(\vb*{\theta}\right)&=- g \vb{D}_\mathrm{s} \vb{\Psi}^T \vb*{m}.\label{eq_g}
\end{align*}

One can notice several points in \eqref{eq:n-link_pilar_D}:
\begin{enumerate}
    \item The first term, of inertial nature, with a global mass matrix which depends on the instantaneous configuration $\vb*{\theta}(t)$. $\vb{M}\left(\vb*{\theta}\right)$ is known as the inertia matrix.
    \item The second term, which depends on both the instantaneous configuration $\vb*{\theta}(t)$ and velocity $\vb*{\dot{\theta}}(t)$, encapsulates the centrifugal and Coriolis-type terms. 
    \item The third term stems from the gravity potential, depending on the instantaneous configuration $\vb*{\theta}(t)$, is of stiffness nature.
\end{enumerate}

From \eqref{eq:n-link_pilar} the acceleration can be computed at each time-step as
\begin{equation*}
    \vb*{\ddot{\theta}} = \vb{M}(\vb*{\theta})^{-1}
    \left( \vb*{\Gamma} - \vb{C}(\vb*{\dot{\theta}},\vb*{\theta}) \vb*{\dot{\theta}} - \vb*{g}(\vb*{\theta}) \right),
\end{equation*}
enabling the numerical time-step integration with any suitable algorithm. Here, a Newmark scheme was used.

\section{Enhancing the dynamical model: a 3-link \emph{casteller} model}\label{sec:3-link_whole}

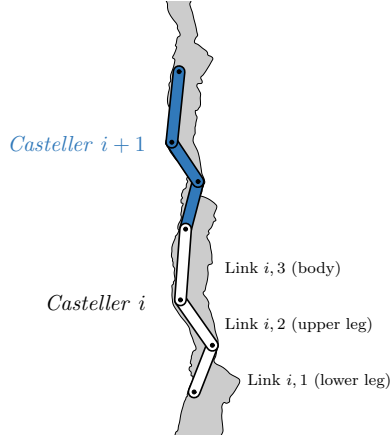
\begin{figure}
\centering	
	\scalebox{0.75}{
    \def\svgwidth{4.0cm}
    \begingroup%
    \setlength{\unitlength}{\svgwidth}%
    \begin{picture}(1,1.91851858)%
        \put(0,0){\includegraphics[width=\unitlength]{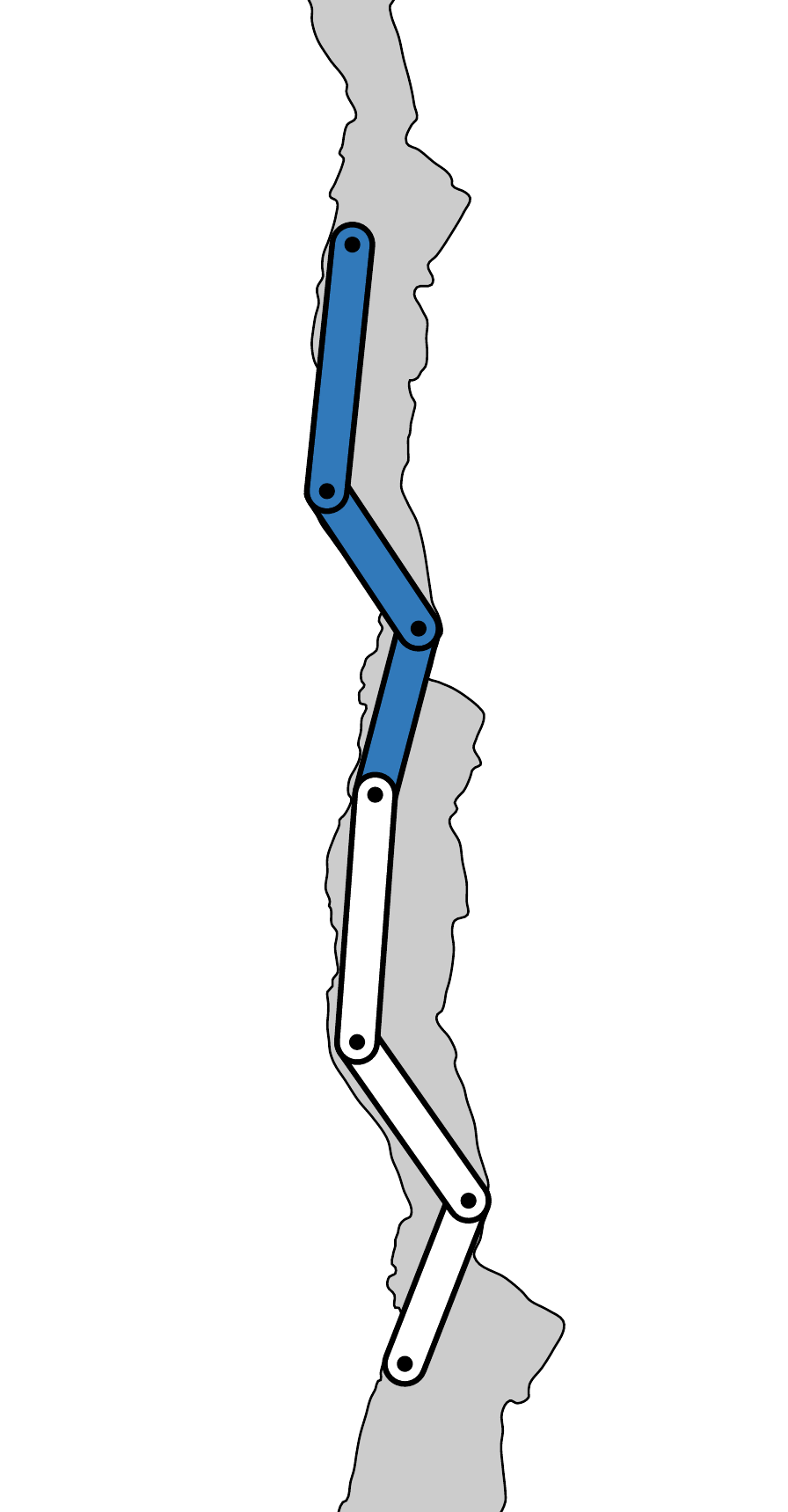}}%
        \put(0.30,0.55){\color[rgb]{0,0,0}\makebox(0,0)[rb]{\emph{Casteller} $i$}}
        \put(0.75,0.20){\color[rgb]{0,0,0}\makebox(0,0)[lb]{\tiny{Link $i,1$ (lower leg)}}}
        \put(0.65,0.45){\color[rgb]{0,0,0}\makebox(0,0)[lb]{\tiny{Link $i,2$ (upper leg)}}}            \put(0.65,0.70){\color[rgb]{0,0,0}\makebox(0,0)[lb]{\tiny{Link $i,3$ (body)}}}    \put(0.30,1.25){\color[rgb]{0.192,0.474,0.729}\makebox(0,0)[rb]{\emph{Casteller} $i+1$}}
    \end{picture}%
    \endgroup%
    }
\caption{Scheme of two \emph{castellers} represented by the 3-link model.}
\label{fig:2castellers3link}
\end{figure}

If we look at a picture of a \emph{pilar}, like in Figure \ref{fig:diagram_pilar}, we observe that the \emph{castellers} do not stand in a vertical position but they bend their knees and hips a little bit (which we refer to as the bent-knee position). This is observed, at a higher or lower degree, in all \emph{castells}, but it is especially noticeable in \emph{pilars}. Therefore, this bending of the joints suggests that modelling each \emph{casteller} as a single rigid link might not be an accurate description of reality and a \emph{casteller} should be composed of several links. Then, one can think that the \emph{pilar} can be directly represented by \eqref{eq:n-link_pilar}, just taking $L\cdot P$ links where $L$ and $P$ are the number of links (per \emph{casteller}) and \emph{castellers}, respectively.

In the Biomechanics literature we can find multiple ways to describe a human body as a set of articulated rigid bars \cite{Kim,Winter}. However, looking back at Figure \ref{fig:diagram_pilar}, we see that, from the structural point of view, each \emph{casteller} can be represented by three main links: one for the lower legs, one for the upper legs and one for the body. See  Figure \ref{fig:2castellers3link}, where \emph{castellers} are $i$-labelled and each link takes values in $j=1,2,3$, representing lower leg, upper leg and body, and their corresponding joints (feet, knees, hip), respectively. Additionally, a closer look shows that there are some complicating issues that arise from considering each \emph{casteller} as a 3-link structure and they will be discussed below.

Along this section, as only a single \emph{casteller} is considered, we omit the sub-index $i$. The aim of this section is to present a simple but more realistic model, including self control, for understanding how a \emph{casteller} behaves when performing a \emph{castell}. This single 3-link \emph{casteller} model is aimed to be a stepping-stone towards modelling a \emph{castell} composed of 3-link \emph{castellers}.

\subsection{A single 3-link \emph{casteller}}\label{sec:3-link}
First, let us obtain the model for a 3-link \emph{casteller}. From the general dynamics obtained in \eqref{eq:n-link_pilar}, and taking $N=3$, a single \emph{casteller} can be represented by the nonlinear equations
\begin{equation}
    \vb{M}\left(\vb*{\theta}\right)\vb*{\ddot{\theta}}+ \vb{C}(\vb*{\dot{\theta}},\vb*{\theta}) \vb*{\dot{\theta}}+
    \vb*{g}\left(\vb*{\theta}\right) = \vb*{\Gamma}_u\label{eq_MechSysLagrangian}
\end{equation}
where $\vb*{\Gamma}_u$ represents now the joint moments of the \emph{casteller} reacting to the external  disturbances, and the inertia, Coriolis and potential related matrices result, respectively, in
\begin{equation*}
\vb{M}(\vb*{\theta})=\begin{pmatrix}
b_{11}&b_{12}\cos(\theta_1-\theta_2)&b_{13}\cos(\theta_1-\theta_3)\\
b_{12}\cos(\theta_1-\theta_2)&b_{22}&b_{23}\cos(\theta_2-\theta_3)\\
b_{13}\cos(\theta_1-\theta_3)&b_{23}\cos(\theta_2-\theta_3)&b_{33}
\end{pmatrix}
\end{equation*}

\begin{equation*}
\vb{C}(\vb*{\dot{\theta}},\vb*{\theta})=\begin{pmatrix}
0&b_{12}\dot{\theta}_2\sin(\theta_1-\theta_2) &b_{13}\dot{\theta}_3\sin(\theta_1-\theta_3)\\
-b_{12}\dot{\theta}_1\sin(\theta_1-\theta_2)&0&b_{23}\dot{\theta}_3\sin(\theta_2-\theta_3)\\
-b_{13}\dot{\theta}_1\sin(\theta_1-\theta_3)&-b_{23}\dot{\theta}_2\sin(\theta_2-\theta_3)&0
\end{pmatrix}
\end{equation*}

\begin{equation*}
\vb*{g}(\vb*{\theta})=-g\begin{pmatrix}
l_1(c_1m_1+m_2+m_3)\sin(\theta_1)\\
l_2(c_2m_2+m_3)\sin(\theta_2)\\
l_3c_3m_3\sin(\theta_3)
\end{pmatrix}
\end{equation*}
with coefficients
\begin{align*}
    b_{11}&=l_1^2(c_1^2m_1+m_2+m_3)+J_1\\
    b_{22}&=l_2^2(c_2^2m_2+m_3)+J_2\\
    b_{33}&=l_3^2c_3^2m_3+J_3\\
    b_{12}&=l_1l_2(c_2m_2+m_3)\\
    b_{13}&=l_1l_3c_3m_3\\
    b_{23}&=l_2l_3c_3m_3.
\end{align*}

The reactions that a \emph{casteller} does in order to avoid the \emph{pilar} from collapsing can be seen from a mathematical point of view as stability and control problems. There exist many control algorithms for systems with the form \eqref{eq_MechSysLagrangian}. Among them, passivity-based controllers ensures asymptotic convergence to the desired trajectory, and have been used in many applications such as robot manipulators \cite{Ortega}. 
Then, the reactions of the \emph{casteller}, can be described as control actions with the form
\begin{equation}\label{eq:controller}
    \vb{\Gamma}_u=\hat{\vb{M}}(\vb*{\theta})(\vb*{\ddot{\theta}_d}-\vb{K}_d(\vb*{\dot{\theta}}-\vb*{\dot{\theta}}_d)-\vb{K}_p(\vb*{\theta}-\vb*{\theta}_d))+\hat{\vb{C}}(\vb*{\dot{\theta}},\vb*{\theta})\vb*{\dot{\theta}}+\hat{\vb*{g}}(\vb*{\theta})
\end{equation}
where $\vb*{\theta}_d$ (and their respective time derivatives) stand for the desired angular positions, velocities and accelerations,  matrices $\vb{K}_d=\text{diag}(K_{d1},K_{d2},K_{d3},),\vb{K}_p=\text{diag}(K_{p1},K_{p2},K_{p3},)$ are the gain matrices that have to be tuned according the desired dynamic response and can be physically related to a damping and elastic effects, respectively, and matrices $\hat{\vb{M}}(\vb*{\theta}),\hat{\vb{C}}(\dot{\vb*{\theta}},\vb*{\theta}),\hat{\vb*{g}}(\vb*{\theta})$ are the estimated ones from the knowledge and perception that each \emph{casteller} have from her/him self. In this case the \emph{casteller} aims for an steady position, so that the desired velocities and accelerations are null,  $\dot{\vb*{\theta}}_d=\ddot{\vb*{\theta}}_d=\vb*{0}$.

The resulting (closed-loop) dynamics of the \emph{casteller} is obtained replacing \eqref{eq:controller} in \eqref{eq_MechSysLagrangian}. In ideal conditions, when estimated matrices in \eqref{eq:controller} match with the actual values in \eqref{eq_MechSysLagrangian}, the closed-loop dynamics simplifies in the decoupled second order dynamics for each angle
\begin{equation*}
    (\vb*{\ddot{\theta}}-\vb*{\ddot{\theta}}_d)+\vb{K}_d(\vb*{\dot{\theta}}-\vb*{\dot{\theta}}_d)+\vb{K}_p(\vb*{\theta}-\vb*{\theta}_d)
     =0, \label{eq_ClosedLoop}
\end{equation*}
where $\vb{K}_d,\vb{K}_p$ unequivocally define the response. Any deviation of the estimated values from the actual ones will affect the performance of the \emph{casteller} and also its stability. 

In this work, as a first approximation and using the previous considerations, the proposed controller has been simplified approximating
$\theta_j\approx 0$ in $\hat{\vb{M}}(\vb*{\theta})$ and $\hat{\vb{C}}(\vb*{\theta})$ (that implies $\hat{\vb{M}}(\vb*{\theta})\approx\vb{I}$ and $\hat{\vb{C}}(\vb*{\dot{\theta}},\vb*{\theta})\approx\vb{0}$), and \eqref{eq:controller} results in
\begin{equation}\label{eq:controllerSimpl}
    \vb{\Gamma}_u=-\vb{K}_d\vb*{\dot{\theta}}-\vb{K}_p(\vb*{\theta}-\vb*{\theta}_d)+\vb*{g}(\vb*{\theta}).
\end{equation}

Notice that the latter simplification is valid for a \emph{casteller} in a complete vertical position. However, as has been previously discussed, we are aiming for representing each \emph{casteller} in a bent-knee position, that means $\vb*\theta_d\neq0$ and, consequently, $\vb{M}(\vb*\theta_d)\neq\vb{I}$. This suggests using \eqref{eq:controller} in a future.

\subsection{Kinematic constraint for the joints}

If we consider a multiple link \emph{casteller}, then we must take into account the different types of joints involved in the problem. We need to distinguish between the internal joints of a \emph{casteller} (namely knees and hip) and the external joint between two different \emph{castellers}, which is the system comprised by the ankles of the top \emph{casteller} and the arms of the bottom \emph{casteller}. One of the key differences between the internal and external joints is that the internal ones show an asymmetric behaviour. Even though that all the joints show kinematic constraints if pushed to extreme enough cases, the internal joints are reasonably close to these constraints. For example, starting from a perfectly vertical position, the hip can only bend forward while the knees can only bend backwards. This limitation can be modelled adding an extra moment vector, $\vb{\Gamma}_\mathrm{kin}$, so that  \eqref{eq_MechSysLagrangian}, results into
\begin{equation*}
    \vb{M}\left(\vb*{\theta}\right)\vb*{\ddot{\theta}}+ \vb{C}(\vb*{\dot{\theta}},\vb*{\theta}) \vb*{\dot{\theta}}+
    \vb*{g}\left(\vb*{\theta}\right) = \vb*{\Gamma}_u+\vb*{\Gamma}_\mathrm{kin}\label{eq_MechSysLagrangianKinConst}
\end{equation*}
where the kinematic constraint can be written as a nonlinear restoring torque defined as
\begin{equation*}
    \Gamma_\mathrm{kin,j} = \max \left(\mp K_\mathrm{kin,j} (\theta_{j+1} - \theta_j - \theta_{\mathrm{lim},j}), 0 \right),
\end{equation*}
where $\theta_j$ and $\theta_{j+1}$ are the angles of the two links connected at the joint, $K_\mathrm{kin}$ is the stiffness of the constraint (which should be very high), $\theta_\mathrm{lim}$ is the limit angle beyond which the joint cannot bend, and the sign depends on which side of the limit angle $\theta_{\mathrm{lim},j}$ is restricted. 

In future work, further differences will arise when the collapse of a \emph{castell} is considered, as the external joints may break but the internal ones cannot.

\subsection{Desired position}\label{sec:desired_pos}
Another aspect to consider when increasing the number of links in a \emph{casteller} is the balance criterion, which is closely related to the controller. When a \emph{casteller} is composed of a single link, it is logical to impose that the balance criterion is to aim for a vertical position. However, when we consider 3-link \emph{castellers} the balance criterion is less obvious. In this section we discuss some of the balance criteria that are taken into account from a single \emph{casteller} point of view in order to maintain the \emph{pilar} to stand, but a further discussion on these mechanisms from a collective point of view of the whole \emph{pilar} (that is, interaction among \emph{castellers} to avoid the \emph{castell} to collapse) will be given in Section \ref{sec:static}. Linking both approaches is, of course, something that should be considered in a future work.

The first balance condition is that the centre of mass of a given \emph{casteller} is vertically aligned with their feet. The physical explanation of this constraint is that a \emph{casteller} aims to reduce the moment applied to the external joint caused by their own weight. Mathematically, this condition can be written as $d = 0$ with
\begin{equation}\label{eq:CoM_horizontal}
    d = -\frac{1}{\sum_{j=1}^3 m_j} \sum_{j=1}^3 m_j \left( c_j l_j \sin \theta_j(t) + \sum_{k=1}^{j-1} l_k \sin \theta_k(t) \right),
\end{equation}
where for each bar $i$ $m_i$ is the mass, $c_i$ is the position of the centre of mass, $l_i$ the length and $\theta_i$ the rotation angle of the bar. This condition is closely related with the \emph{mass center eccentricity} described in Section \ref{sec:static}. Actually, formula \eqref{eq:CoM_horizontal} will be generalized in formula \eqref{eq:di} when taking into account the displacement of the \emph{castellers} below (see Section \ref{sec:static} for more details). 

The second balance condition is that the shoulders of a given \emph{casteller} are vertically aligned with their feet too. The physical explanation of this constraint is that a \emph{casteller} aims to align the force received from the \emph{castellers} above with their feet to reduce the moment generated. Mathematically, this condition can be written as $w = 0$ with
\begin{equation}\label{eq:Shoulders}
    w = -\sum_{j=1}^3 l_j \sin \theta_j(t).
\end{equation}

Finally, the last balance condition is that the centre of mass of a given \emph{casteller} is at a given height. Mathematically, this condition can be written as $h = h_d$ with
\begin{equation}\label{eq:CoM_vertical}
    h = \frac{1}{\sum_{j=1}^3 m_j}\sum_{j=1}^3 m_j \left( c_j l_j \cos \theta_j(t) + \sum_{k=1}^{j-1} l_k \cos \theta_k(t) \right),
\end{equation}
and $h_d$ being a desired position. The physical interpretation of this last condition is less clear but it represents the fact that certain bent-knee positions are more comfortable to sustain for long times than others. The comfortable position is described in this case by $h_d$. This condition is closely related with the \emph{folding index} described in Section \ref{sec:static}.

Defining the position of a \emph{casteller} in a compact way as
\begin{equation}
\vb*{\xi}(\vb*{\theta})=\begin{pmatrix}d(\vb*{\theta})\\w(\vb*{\theta})\\h(\vb*{\theta})\end{pmatrix}\label{eq_xi}
\end{equation}
the new variable $\vb*{\xi}$ provides the necessary change of coordinates to obtain the desired steady angles as
\begin{equation}
\vb*{\theta}_d=\vb*{\xi}^{-1}(0,0,h_d).\label{eq_thetaDesired}
\end{equation}
A further analysis on the existence of \eqref{eq_thetaDesired} would be necessary to ensure a set of admissible target angles without violating physical constraints. Also, the results of Section \ref{sec:static} could help in this analysis.

The desired position in terms of angles \eqref{eq_thetaDesired} can be directly used in \eqref{eq:controller} and the \emph{casteller} will reach the desired position. However, this control strategy does not coordinate the angles in order to reach the target position, which makes it not very realistic.
For a more accurate model, it is necessary to provide reactions of the \emph{casteller} able to look for the  balance conditions defined as \eqref{eq:CoM_horizontal}, \eqref{eq:Shoulders} and \eqref{eq:CoM_vertical}. As we said, that is something that will be considered in Section \ref{sec:static}, but not from a control point of view. Hence, an interesting future work would be to find a more realistic controller able to simulate better the coordinated reactions of the \emph{casteller} at each joint by analysing the dynamics of $\vb*{\xi}$ in \eqref{eq_xi}.

\subsection{Delay-differential balance model}
Another effect that could be taken into account is the delay in the human-induced restoring moments (see \cite{Chadges_etal,Kow_etal}). The delay represents the reaction time of a \emph{casteller} when trying to recover the balance. In this case, we can write the general form of the controller as 
\begin{equation}\label{eq:controller_delay}
\vb{\Gamma}(t)=f(\dot{\vb*{\theta}}(t-\tau),\vb*{\theta}(t-\tau)),  
\end{equation}
where $f$ is the controller function, which could be set to have a similar form to \eqref{eq:controller} or \eqref{eq:controllerSimpl}, and $\tau$ is the delay parameter. The interpretation of \eqref{eq:controller_delay} is that the reaction of the \emph{casteller} at time $t$ is based on the position at time $t - \tau$ where the delay parameter represents the reaction time. Note that in \eqref{eq:controller_delay} we assume that the delay parameter is the same for all the joints, however it could be reformulated to account for different delay parameters and, in the future, for different \emph{castellers} (according to their biomechanical characteristics and expertise). In addition, we could assume that the delay parameter changes in time, as \emph{castellers} become progressively more fatigued.

Preliminary simulations of an inverted pendulum modelling a pillar (with \emph{castellers} represented by a single link model) have shown that delays associated to neuro-muscular behaviours, reaction times and fatigue, could dramatically compromise the stability of the structure. See videos in \cite{simulations}.


\subsection{Benefits of the bent-knee position}
One of the questions that came up when posing the 3-link \emph{casteller} model was what is the advantage of the bent-knee position. To study this, we studied a system composed by two links with an asymmetric joint. The equations for this system are the subset of equations for $\theta_1$ and $\theta_2$ in Section \ref{sec:3-link}, setting $m_3 = 0$. 

In this example we use the PD-like controller defined in \eqref{eq:controllerSimpl} without any delay effects. The desired position is $\vb*{\ddot{\theta}}_d = \vb*{\dot{\theta}}_d = \vb*{0}$ with a $\vb*{\theta}_d$ to be specified. We impose that $\vb*{\theta}_d$ must satisfy the condition that the centre of mass is aligned with the feet (similar to $d = 0$ in \eqref{eq:CoM_horizontal}), which leaves one degree of freedom in our problem. For the simulation we start at $\vb*{\theta} = \vb*{\theta}_d$ with initial angular velocities $\theta_1 = 0$ rad s$^{-1}$ and $\theta_2 = 10$ rad s$^{-1}$. The rest of the model parameters are set to
\begin{equation*}
    \begin{aligned}
    l_1 &= 1.80, & c_1 &= 0.56, & m_1 &= 100, & J_1 &= 27.00, & K_{p1} &= 10000, & K_{d1} &= 100,\\
    l_2 &= 2.25, & c_2 &= 0.56, & m_2 &= 125, & J_2 &= 52.73, & K_{p2} &= 12500, & K_{d2} &= 125,
    \end{aligned}
\end{equation*}
with $g = 9.81$.

To test the benefits of the bent-knee position, we run the simulation for three different desired positions: standing ($\theta_{d,1} = 0^\circ$), small bending ($\theta_{d,1} = -10^\circ$) and large bending ($\theta_{d,1} = -20^\circ$). The desired position of $\theta_2$, defined as $\theta_{d,2}$, is chosen to satisfy that the centre of mass is aligned with the feet.

\begin{figure}[htpb]
\centering
\includegraphics[width=\textwidth]{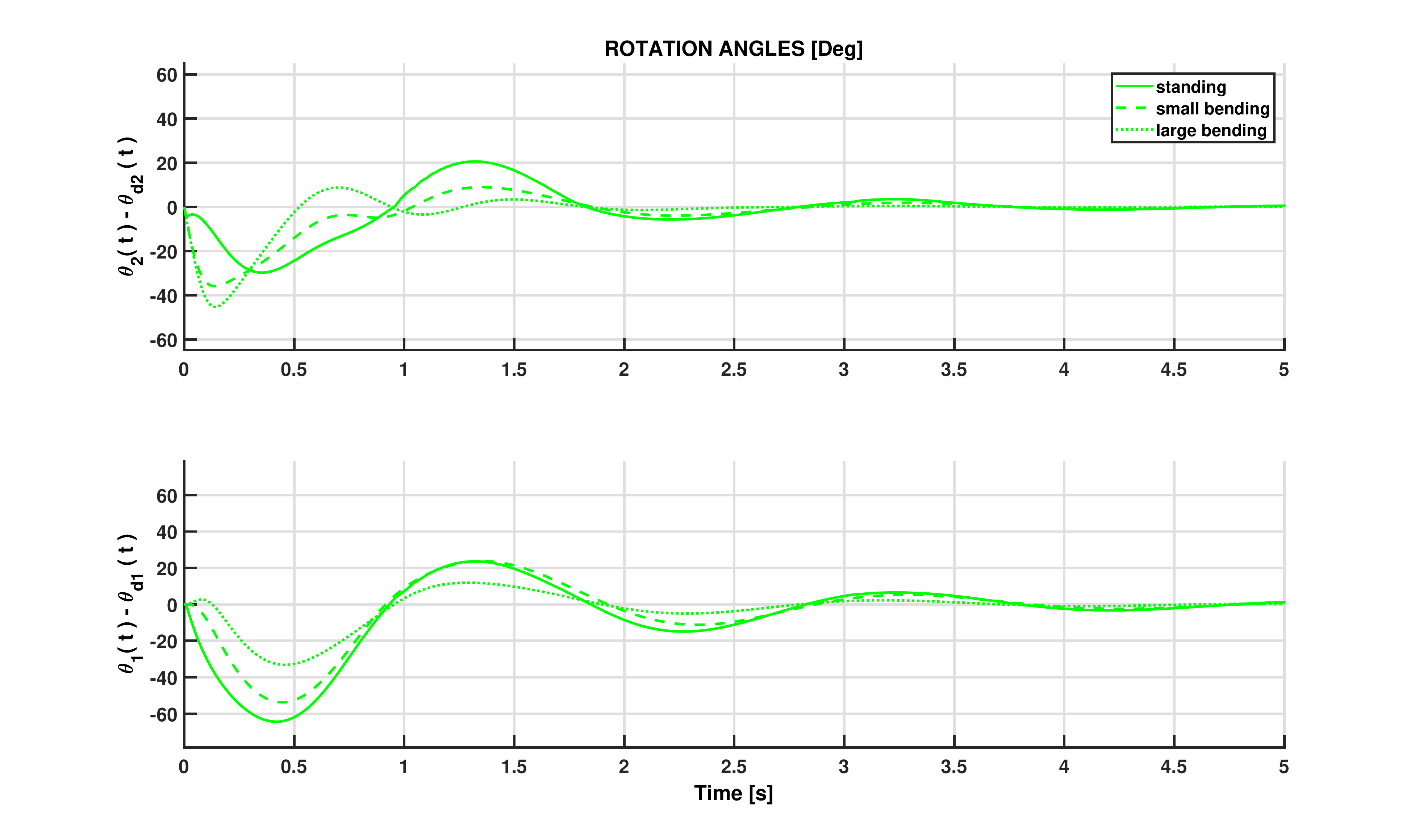}
\caption{Values for the difference between angles $\theta_1$ and $\theta_2$ and their corresponding equilibrium position as a function of time for the three bent-knee positions. }
\label{fig:bent-knee_angle}
\end{figure}

\begin{figure}[htpb]
\centering
\includegraphics[width=\textwidth]{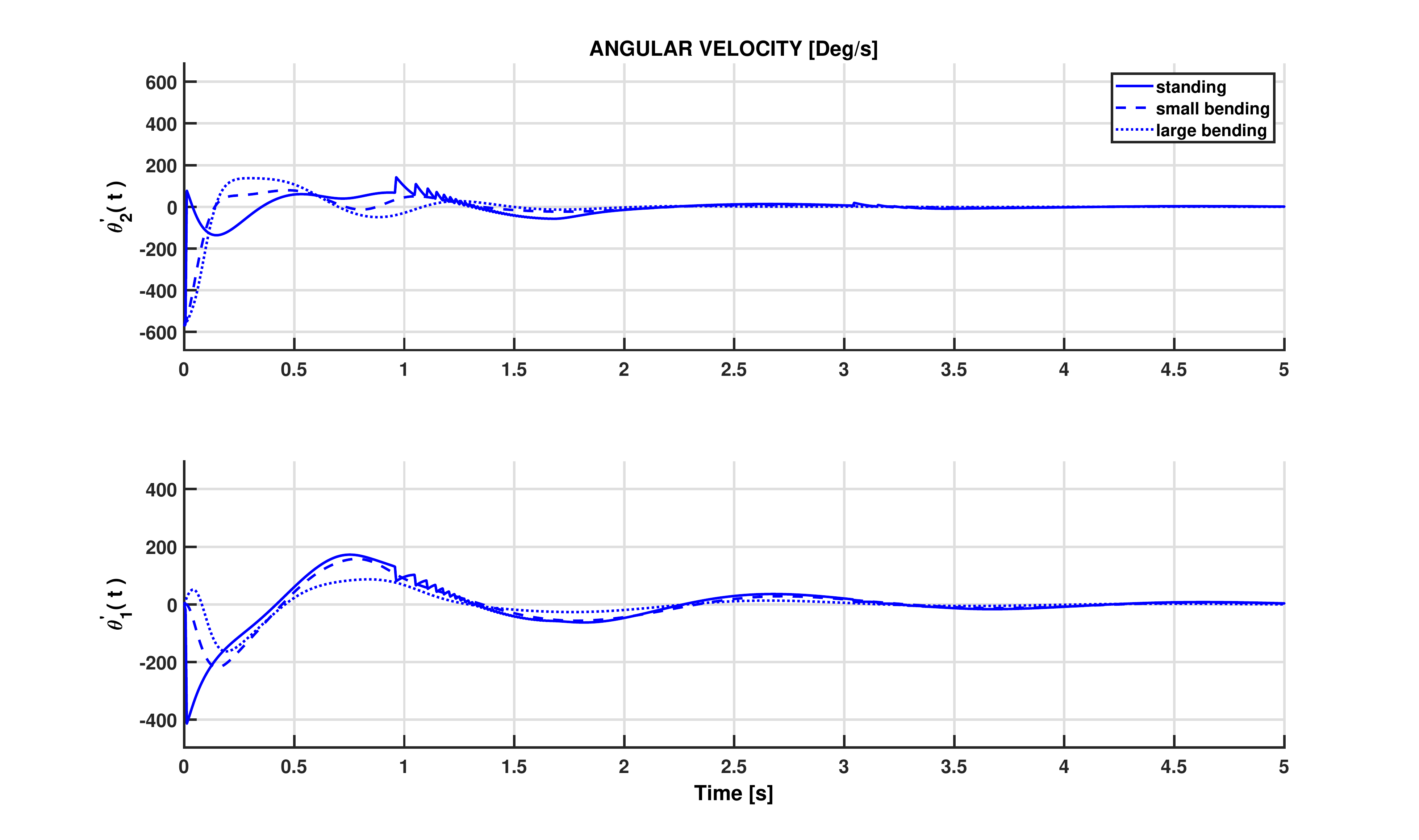}
\caption{Values for the angular velocities $\dot{\theta}_1$ and $\dot{\theta}_2$ as a function of time for the three bent-knee positions.}
\label{fig:bent-knee_angular_velocity}
\end{figure}

\begin{figure}[h!]
\centering
\includegraphics[width=\textwidth]{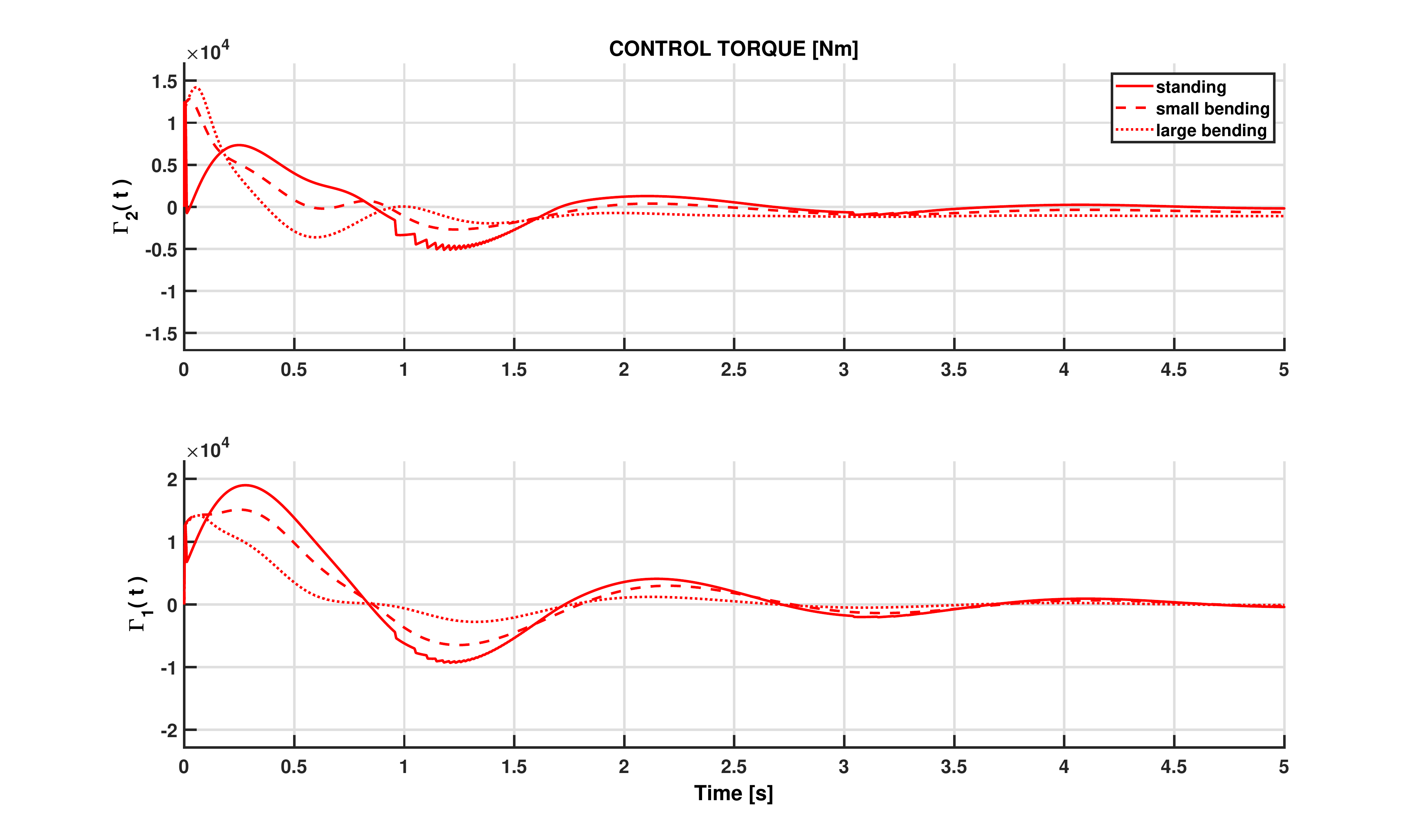}
\caption{Values for the torques $\Gamma_1$ and $\Gamma_2$ as a function of time for the three bent-knee positions.}
\label{fig:bent-knee_torque}
\end{figure}

The results of the simulations are shown in Figures \ref{fig:bent-knee_angle}-\ref{fig:bent-knee_torque}. We observe that, even though the control torques are initially larger, the system returns faster to the equilibrium positions the more bent these are. We suspect that this is because, in those cases, the kinematic constraint does not act which makes it easier for \emph{castellers} to recover balance.

\section{A static balance point of view}\label{sec:static}

\begin{figure}[htpb]
\centering
	\scalebox{0.75}{
    \def\svgwidth{2.5cm}
    \begingroup%
    \setlength{\unitlength}{\svgwidth}%
    \begin{picture}(1,3.55624871)%
        \put(0,0){\includegraphics[width=\unitlength]{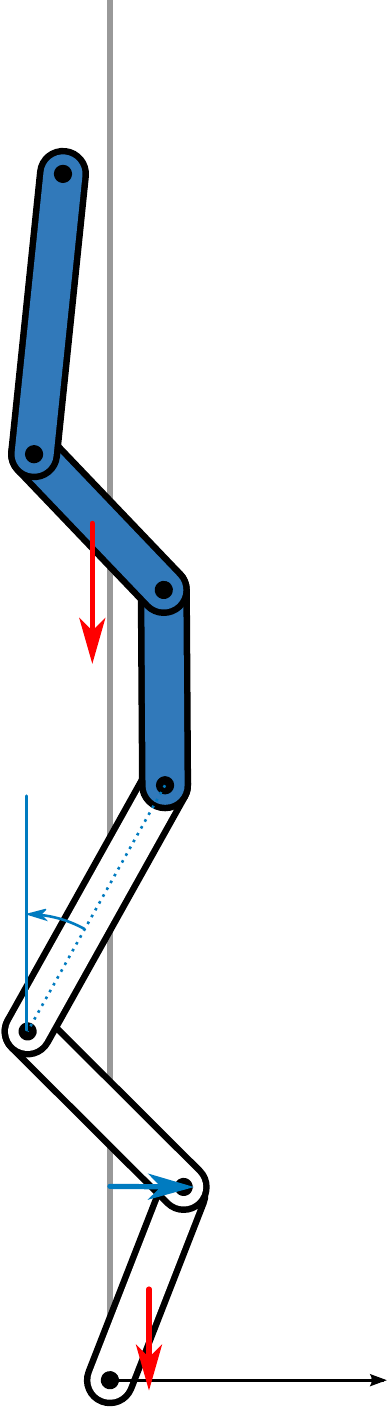}}%
        \put(1.00,-0.1){\color[rgb]{0,0,0}\makebox(0,0)[rb]{$d$}}        \put(0.70,0.90){\color[rgb]{0,0,0}\makebox(0,0)[lb]{\emph{Casteller} $1$}}
        \put(0.70,2.40){\color[rgb]{0.192,0.474,0.729}\makebox(0,0)[lb]{\emph{Casteller} $2$}}
        \put(0.25,0.45){\color[rgb]{0.192,0.474,0.729}\makebox(0,0)[rb]{$d_{1,2}$}}        \put(-0.05,1.20){\color[rgb]{0.192,0.474,0.729}\makebox(0,0)[rb]{$\theta_{1,3}$}}
        \put(0.450,0.10){\color[rgb]{1,0.0,0.0}\makebox(0,0)[lb]{$m_{1,1}g$}}        \put(0.20,2.00){\color[rgb]{1,0.0,0.0}\makebox(0,0)[rb]{$m_{2,2}g$}}
        \put(1.21,1.50){\color[rgb]{0,0,0}\makebox(0,0)[rb]{\tiny{(feet, $j=1$)}}}        \put(1.22,2.05){\color[rgb]{0,0,0}\makebox(0,0)[rb]{\tiny{(knee, $j=2$)}}}        \put(-0.10,2.40){\color[rgb]{0,0,0}\makebox(0,0)[rb]{\tiny{(hip, $j=3$)}}}
    \end{picture}%
    \endgroup%
    }

\caption{Sketch of a static model configuration, that complements notation given in Figure \ref{fig:2barsscheme}.}\label{Croquis}
\end{figure}

The approach in this section is slightly different from the followed in the previous ones. We here consider a given and fixed configuration for a \emph{pilar} and compute the torque distributions caused by the eccentricity of self-weighted loads in a non-perfectly vertical pillar. This is done considering a static balance of forces and moments, where the \emph{castellers} are modelled again as a set of three links (see Figure \ref{Croquis} and Section \ref{sec:3-link_whole}), statically interacting between them and, of course, the rest of the \emph{castellers} of the \emph{tronc}.

We start by recalling some notation given in the previous sections: each of the links (and corresponding joints) of a \emph{pilar} is labelled $i,j$, where $i= 1 \ldots N$ represents the \emph{casteller} ($1$ corresponding to the bottom one, $N$ corresponding to the top one, see Section \ref{sec:dynam}), and $j=1,2,3$ represents the bars (and the joints) of each \emph{casteller} (see Section \ref{sec:3-link_whole}). See Figure \ref{Croquis} as example.


In the whole section, we are going to assume that no \emph{castellers} are ascending or descending, that is no \emph{castellers} are at the back of the other ones.
The given configuration is parametrized by the bar rotations $\theta_{i,j}$ (measured counterclockwise from the vertical; the angle $\theta_{1,3}$ in Figure \ref{Croquis}  is negative). Finally, a set of biomechanical data will be also provided: $c_{i,j}, l_{i,j}, m_{i,j}$ represent the relative location of the mass center (i.e. $c_{i,j}=0.5$ would mean a center of mass located at the midpoint of the bar), the length and the mass of the $ij$-bar. 



In order to compute the torque distribution, we first need to compute some auxiliary variables. Using simple trigonometry it is easy to see that the horizontal displacement of the $ij$-joint, $d_{i,j}$ (with respect to the feet of \emph{casteller} 1 and the usual sign convention, i.e rightward displacements are positive), can be computed as
\begin{eqnarray*}
  d_{1,1} &=& 0 \\
  d_{i,1} &=& d_{i-1,3} - l_{i-1,3}\,\sin{(\theta_{i-1,3})},\ \ \textrm{for } 1<i\leq N\\
  d_{i,j} &=& d_{i,j-1} - l_{i,j-1}\,\sin{(\theta_{i,j-1})},\ \ \textrm{for } 1\leq i\leq N, \ j=2,3.
\end{eqnarray*}


From $d_{i,j}$ we can compute the horizontal displacement of the $ij$-bar mass center, $a_{i,j}$:
\begin{eqnarray*}
   a_{i,j} &=& (1-c_{i,j})d_{i,j} + c_{i,j}\, d_{i,j+1},\ \ \textrm{for } 1\leq i\leq N, \, j=1,2 \\
   a_{i,3} &=& (1-c_{i,3})d_{i,3} + c_{i,3}\, d_{i+1,1},\ \ \textrm{for } 1\leq i< N\\
   a_{N,3} &=& d_{N,3} - c_{N,3}\,  l_{N,3}\,\sin{(\theta_{N,3})}.
\end{eqnarray*}

\vspace{0.2cm}
\noindent and also the displacement of the $i$-\emph{casteller} mass center is:
\begin{equation}\label{eq:di}
   d_i = \dfrac{\sum_{j=1}^3 m_{i,j} a_{i,j} }{m_i},\ \ i=1,\ldots, N 
\end{equation} 
where $m_i= \sum_{j=1}^3 m_{i,j}$ stands for the mass of $i$-\emph{casteller}.

With all these ingredients, the torque $\Gamma_{i,j}$ supported at the $ij$-joint can be computed as
\begin{equation}\label{eq:torque}
  \Gamma_{i,j} = \sum_{k=j}^3 m_{i,k} g \left( a_{i,k}- d_{i,j} \right) + \sum_{k=i+1}^N m_k g \left( d_k-d_{i,j} \right),\ \ i=1,\ldots, N,\ j=1,2,3
\end{equation}  
where $g$ stands for the gravitational acceleration. The first summation term in the right hand side corresponds to the torque due to self-weight of \emph{casteller} $i$ above joint $j$, while the second one corresponds to the torque from the \emph{castellers} above. We recall that we are assuming that no \emph{castellers} are descending, that is, no \emph{castellers} are at the back of the \emph{casteller} $i$. With our sign conventions, a negative supported torque indicates that the pillar is leaning forward (to the right). 

\vspace{0.2cm}
\noindent\textbf{Quality indicators.} The previous computations can be used to define three heuristic quality indicators for a \emph{castell}, aimed at determining whether a given angle configuration is feasible to be maintained or, on the contrary, will collapse. All of them can be computed at each joint or averaged over each floor or the whole They complement the balance criterion given in Section \ref{sec:desired_pos}, where no interactions with the rest of the floors are considered.

These indicators are the \emph{mass center eccentricity} with respect to the vertical, the accumulated horizontal displacement or \emph{folding index}, and the \emph{suffering index}, whose objective is evaluating how much the joints are suffering.

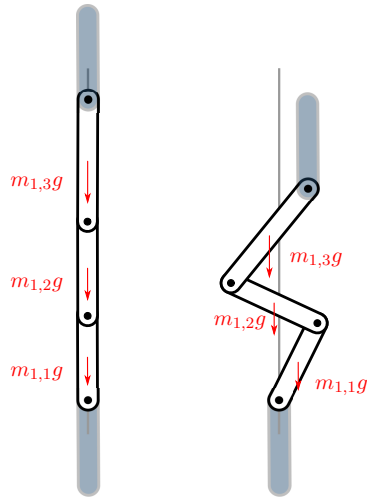
\begin{figure}[htpb]
\centering
	\scalebox{0.75}{
    \def\svgwidth{4.5cm}
    \begingroup%
    \setlength{\unitlength}{\svgwidth}%
    \begin{picture}(1,1.94905161)%
        \put(0,0){\includegraphics[width=\unitlength]{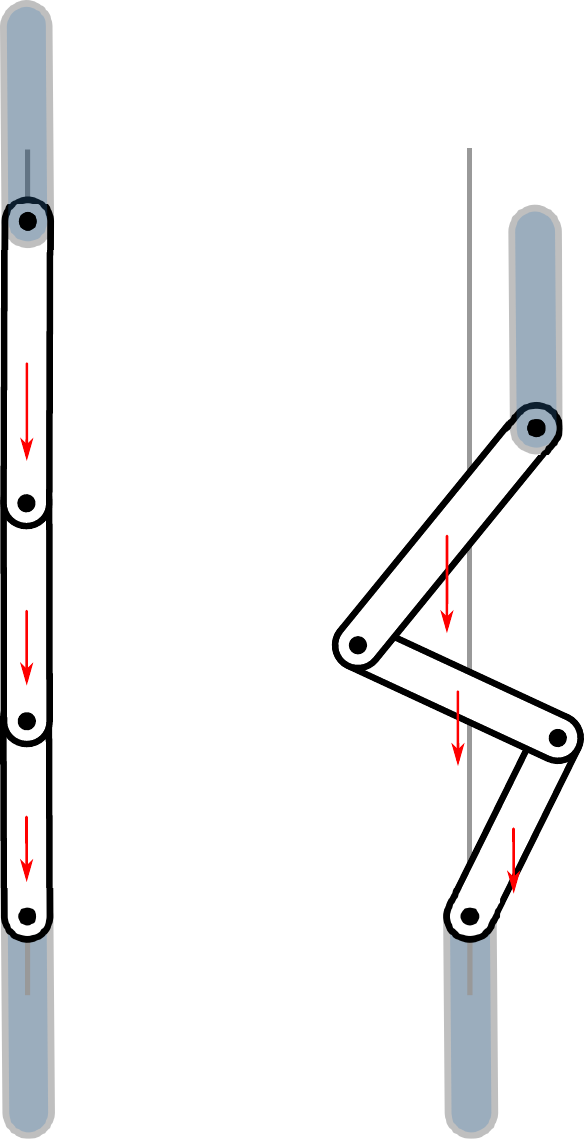}}%
        \put(-0.05,1.20){\color[rgb]{1,0.0,0.0}\makebox(0,0)[rb]{$m_{1,3}g$}}        \put(-0.05,0.8){\color[rgb]{1,0.0,0.0}\makebox(0,0)[rb]{$m_{1,2}g$}}
        \put(0.0,0.45){\color[rgb]{1,0.0,0.0}\makebox(-0.05,0)[rb]{$m_{1,1}g$}}
        \put(1.05,0.90){\color[rgb]{1,0.0,0.0}\makebox(0,0)[rb]{$m_{1,3}g$}}        \put(0.75,0.65){\color[rgb]{1,0.0,0.0}\makebox(0,0)[rb]{$m_{1,2}g$}}
        \put(1.15,0.40){\color[rgb]{1,0.0,0.0}\makebox(0,0)[rb]{$m_{1,1}g$}}
    \end{picture}%
    \endgroup%
    }
\caption{Mass center eccentricity equals $0 $ for a perfectly vertical pillar (left) and a folded one (right). But folding index only equals $0$ for the perfectly vertical pillar (left).}\label{EccFolding}
\end{figure}

The \emph{mass center eccentricity} is the horizontal distance of the load above a given joint to the origin of coordinates (that is, the feet of the bottom-most \emph{casteller}). 
We can compute this eccentricity in each floor or \emph{casteller} ($E_i$) or for the whole \emph{castell} ($E_T$):
\begin{equation}\label{eq:eccentricity}
  E_i = \dfrac{\Gamma_{i,1}}{\sum_{j=i}^N m_j\,g},\ \ \   E_T = \dfrac{\Gamma_{1,1}}{M g}
\end{equation}
(here $M=\sum_{i=1}^N m_i$ represents the total mass of the \emph{castell}).
It is a mechanical indicator of the average verticality of a \emph{castell} in the following sense: a perfectly vertical pillar has $E_T=0$, but a pillar with significant rotations $\theta_{i,j}$ may also have $E_T=0$ if positive and negative rotations compensate each other (see Figure \ref{EccFolding}).
In this sense, it can be seen as an external indicator of the quality of a \emph{castell}: if it is larger than a certain threshold, the \emph{castell} bends to much with respect to the vertical and, hence, it is more likely to collapse. 

The \emph{folding index} is defined as the accumulated horizontal displacement. Again, it can be computed at each floor $\Delta_i$ or for the whole \emph{castell} $\Delta_T$:
\begin{equation}\label{eq:folding}
  \Delta_i =\sum_{j=1}^3 | d_{i,j}|,\ \ \   \Delta_T =\sum_{i=1}^N \Delta_i.
\end{equation}
This index tells between structures with similar global deviation but its larger in those where the floors are more \emph{folded}, that is, for the ones at which \emph{castellers} are more bent. This is due to the fact that the absolute value precisely prevents for compensation when $d_{i,j}$ have different signs. Note that only a perfectly vertical pillar has $\Delta_T=0$ (see Figure \ref{EccFolding}). In this sense, we can think it also as an external, aesthetic and visual indicator of the quality of a \emph{castell}. But we can also think it as a measure of the ability that a pillar has to react to perturbations, because a too large value of this index reveals a more fatiguing position for the involved \emph{castellers} (and hence, less ability to react), but also a too small one may also mean that \emph{castellers} are less able to react due to biomechanical reasons.


Finally, we define the \emph{suffering index} as the ratio between the computed needed torque in each joint and the maximum torque this same joint can offer:
\begin{equation*}
  S_{i,j}= \dfrac{\Gamma_{i,j}}{\Gamma_{i,j}^{max}}.
\end{equation*}
This indicator can be computed also in each floor, $\Gamma_i^{rel}$, or for the the whole \emph{castell}, $S_T$:
\begin{equation*}
  S_{i}= \dfrac{\sum_{j=1}^3\Gamma_{i,j}}{\sum_{j=1}^3\Gamma_{i,j}^{max}},\ \ \   S_T = \dfrac{\sum_{i=1}^N\sum_{j=1}^3\Gamma_{i,j}}{\sum_{i=1}^N\sum_{j=1}^3\Gamma_{i,j}^{max}}.
\end{equation*}
It aims at quantifying how much the joints of the \emph{castellers} are suffering in a certain position: if it is larger than a certain threshold, the \emph{castell} will collapse, as that joint will not be physically able to maintain that torque. That would mean that with this indicator we could also get to know which particular joints are the ones that may have caused a \emph{castell} to collapse. In this sense, we can think of the suffering index as an internal and mechanical indicator. It is worth noticing that the maximum allowed torques depend on biomechanical factors but also, of course, on the quality of a \emph{casteller}, and that may decrease with time. In this sense it will be interesting to include in a future work the fatigue that the \emph{castellers} are experiencing, as it may change the maximum moment allowed at certain joints.

\vspace{0.2cm}
\noindent\textbf{Numerical results.} To see how these quality indicators relate with the external and internal feelings of the quality of a \emph{castell}, we evaluate them for two specific \emph{pilars}, a successful one and one that ends up collapsing. Both of them are of the same type, \emph{pilar de 8 amb folre i manilles (p8fm)}, which turns out to be the most difficult type of pillars that can be done.

\begin{figure}[htpb!]
\centering
\includegraphics[width=3cm,height=5.5cm]{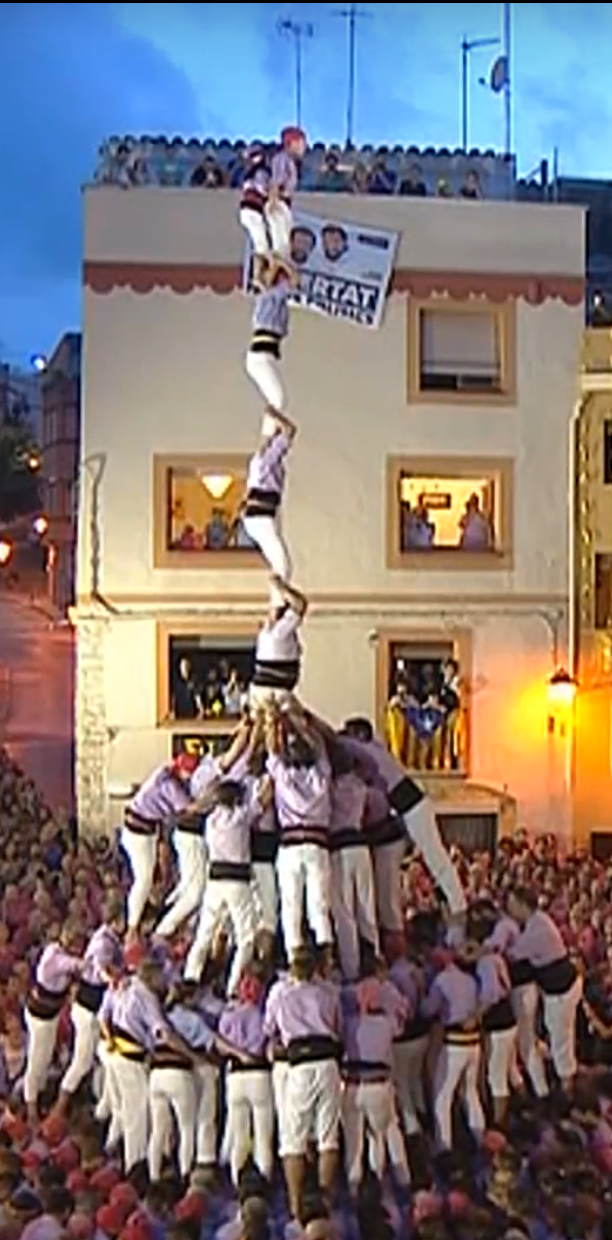}\hspace{1.5cm}\includegraphics[width=8cm]{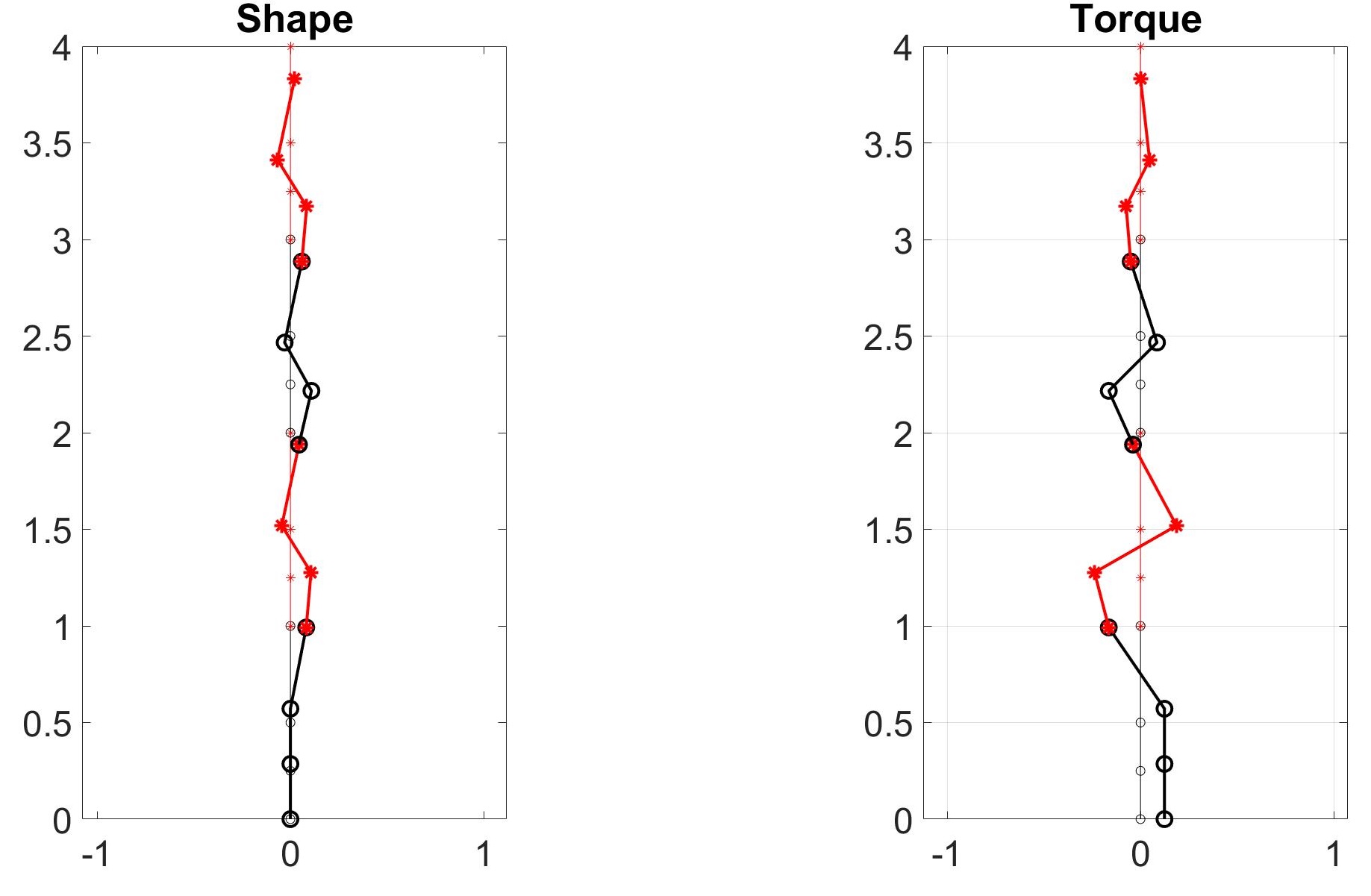}
\caption{Joint positions and received or supported torques in the \emph{p8fm} of Colla Jove Xiquets de Tarragona (El Catllar, 2018) (frame taken from \url{https://youtu.be/rCxT1rFLj9I}, minute 3:35 aprox.). See the corresponding quality indicators on Table \ref{taula}. }\label{JXT}
\end{figure}

\begin{figure}[htpb!]
\centering
\includegraphics[width=3cm,height=5.3cm]{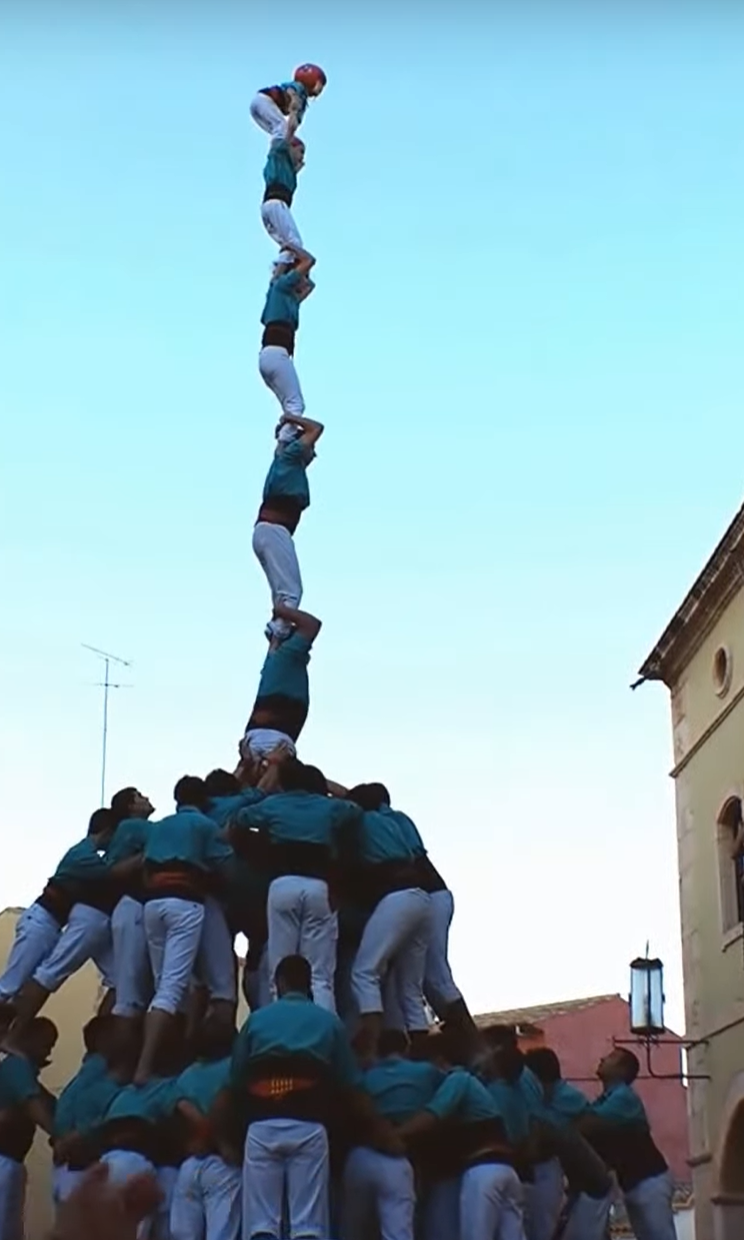}\hspace{1.5cm}\includegraphics[width=8cm]{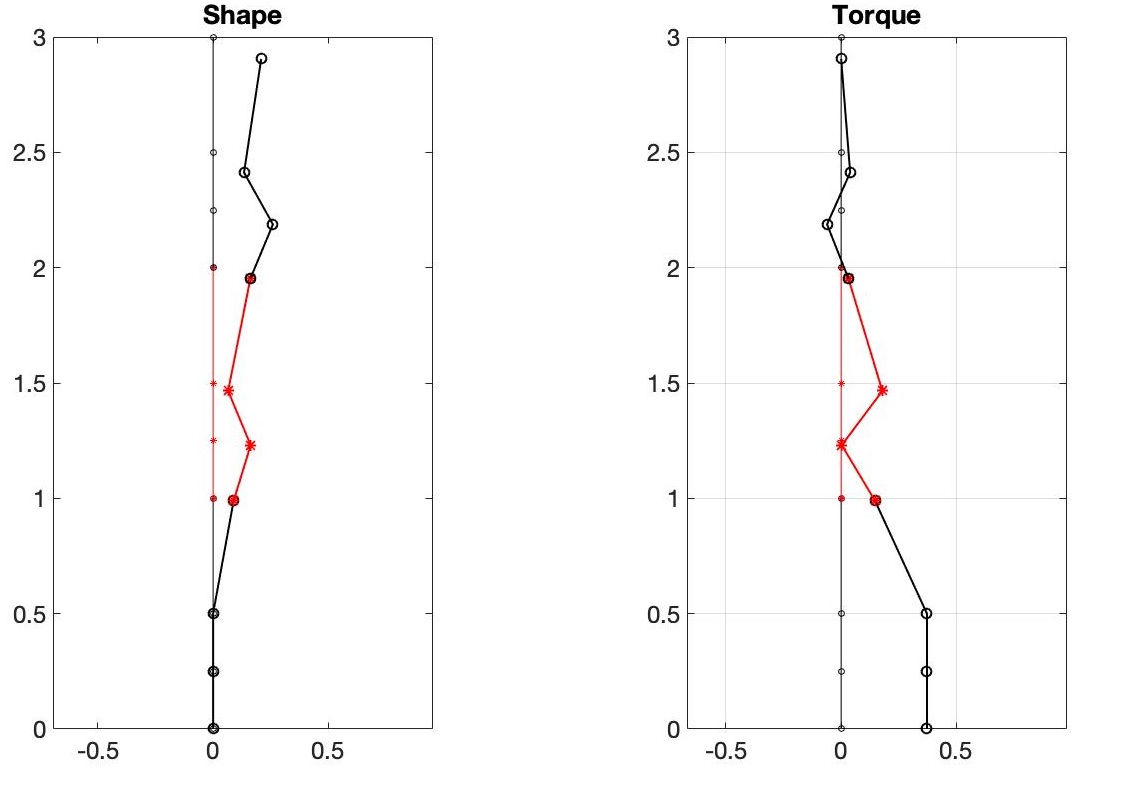}\\
\vspace{0.2cm}
\includegraphics[width=3cm,height=5.3cm]{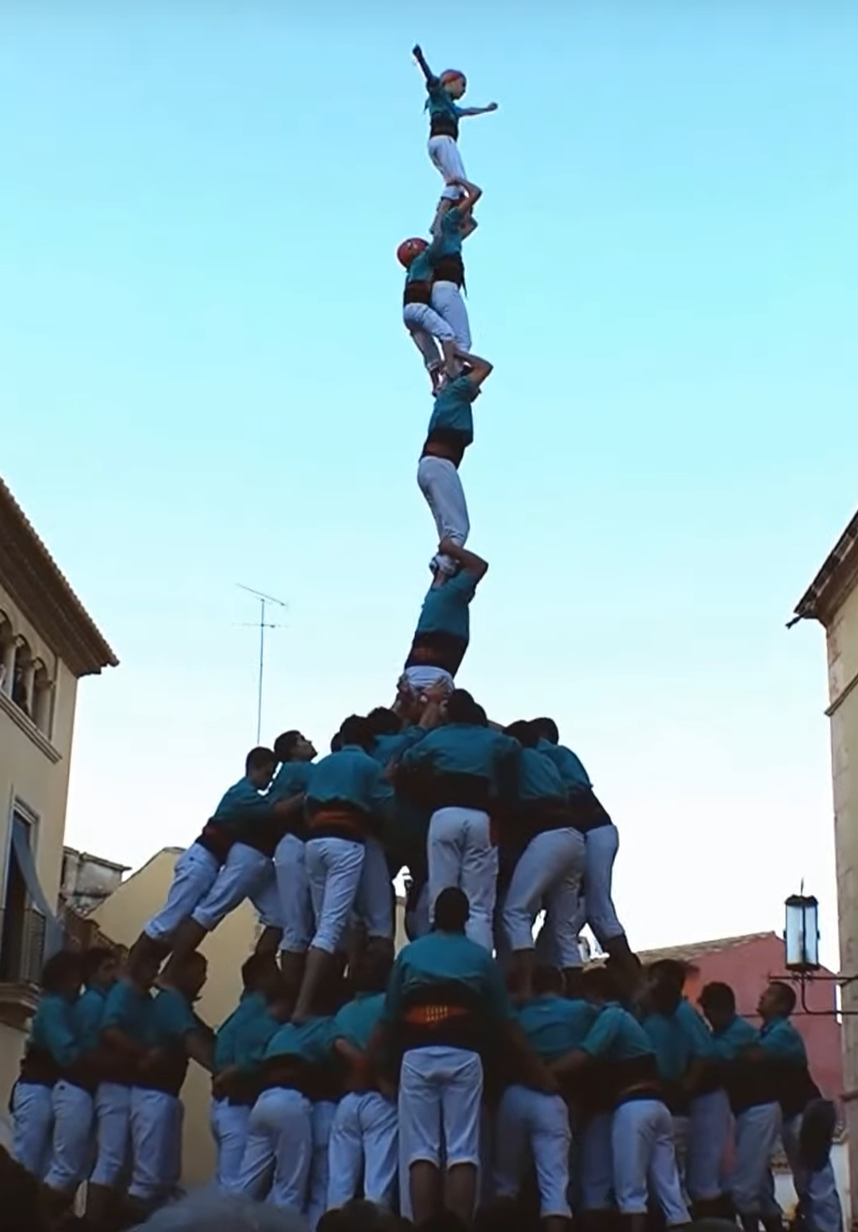}\hspace{1.5cm}\includegraphics[width=8cm]{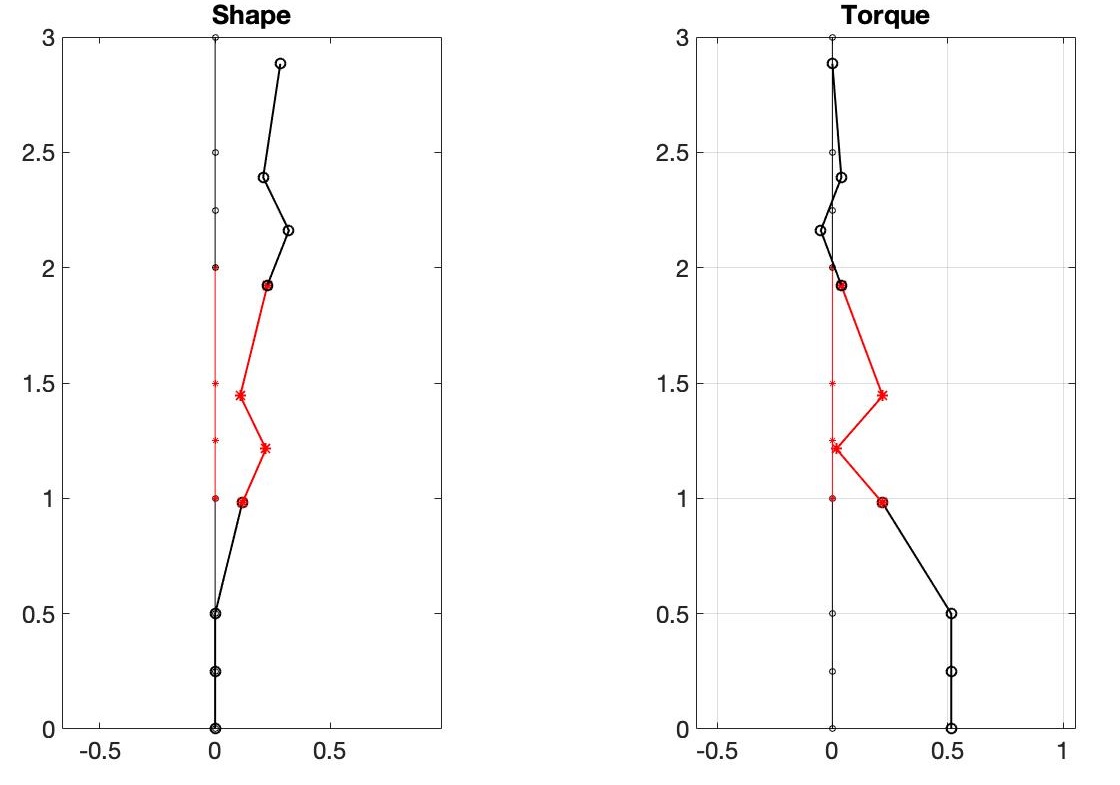}\\
\vspace{0.2cm}
\includegraphics[width=3cm,height=5.3cm]{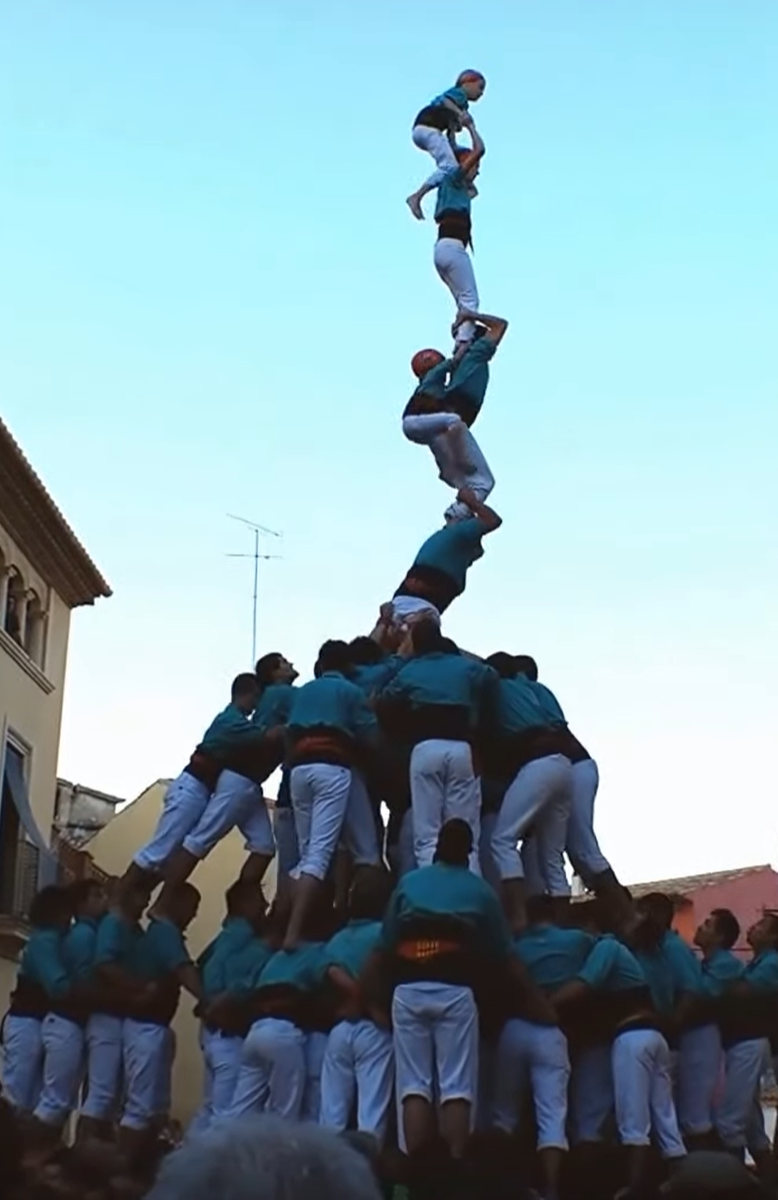}\hspace{1.5cm}\includegraphics[width=8cm]{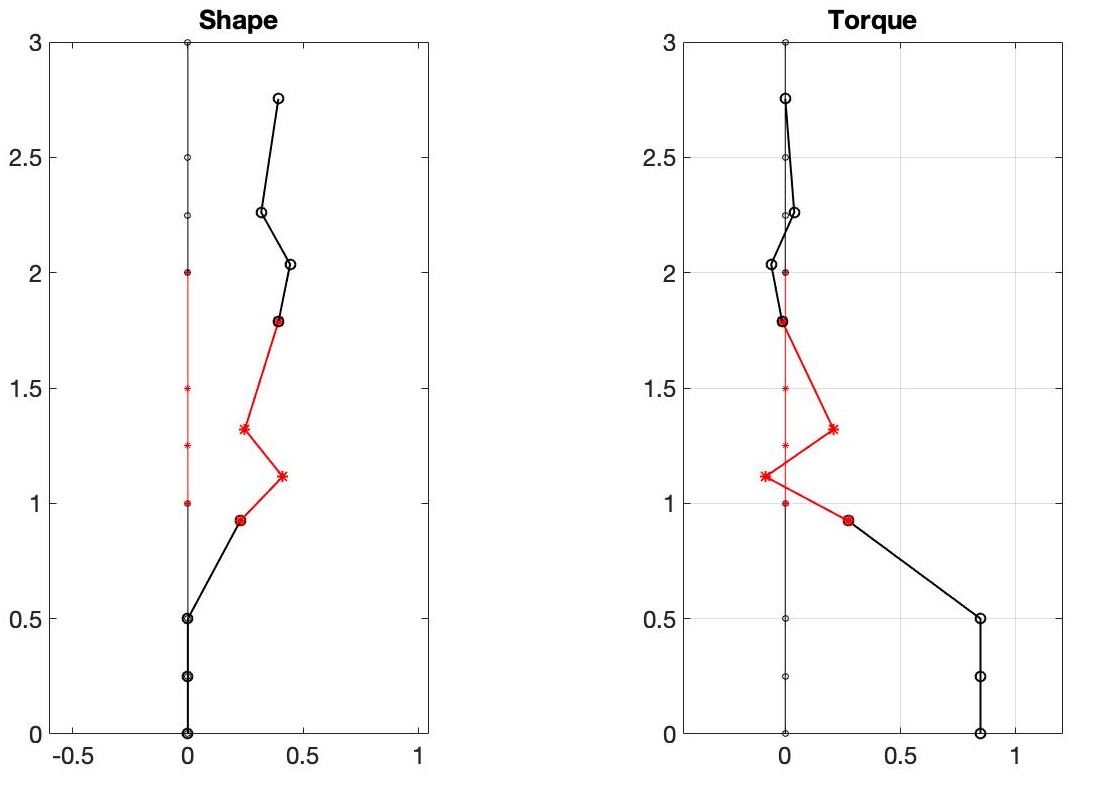}
\caption{Joint positions and received or supported torques in the \emph{p8fm} of  Castellers de Vilafranca (Altafulla, 2014) (frames taken from \url{https://youtu.be/6o7Eaka7vcw}, from minute 3:15). See the corresponding quality indicators on Table \ref{taula}.}\label{Vilaf}
\end{figure}

The first one corresponds to the pillar that the \emph{Colla Jove Xiquets de Tarragona (CJXT)} performed in the \emph{Diada d'El Catllar} (2018) (see left picture of Figure \ref{JXT}). And the one that ends up collapsing corresponds to the \emph{Castellers de Vilafranca (CdV)}, and was performed in the \emph{Diada d'Altafulla} (2014) (see pictures on the left column of Figure \ref{Vilaf}). In both cases, we consider fixed frames of each pillar and measure the angle configuration of each one (one frame for the first pillar, but three for the second one, in order to compute all the variables at different stages of the failure). The corresponding angle configuration can be seen in Table \ref{angles}.
In a first approach, we estimate the following biomechanical parameters as
\begin{eqnarray}\label{biomech}
&&  c_{i,1}= 0.5,\ \ c_{i,2}= 0.5, \ \ c_{i,3}=0.75\\ 
&&  l_{i,1}= \dfrac{2}{7}\,L,\ \ l_{i,2}= \dfrac{2}{7}\,L, \ \ l_{i,3}=\dfrac{3}{7}\,L, \ \ \textrm{with }L=1 \nonumber \\
&&  m_{i,1}= 0.1,\ \ m_{i,2}= 0.2, \ \ m_{i,3}=0.7, \ \ \textrm{with }M=1 \nonumber
\end{eqnarray}
for all $i=1,\ldots N$, and $L$ standing for the height of a \emph{casteller} (from feet to shoulders), that we take equal to 1 for simplicity. Observe that in this section, and again for simplicity, we are assuming that the \emph{pilar} is built with identical \emph{castellers} in each floor. The appropriate values of these parameters is, of course, something that should be considered with more detail in a future work, as well as their heterogeneity among different \emph{castellers}. Also, in these examples we are only considering the $N=4$ \emph{castellers} that go from levels four to seven, in spite that the \emph{pilar} has eight levels in total (that is, $i=1,2,3,4$ correspond in this examples to the \emph{quart}, \emph{quint}, \emph{sis\`e} and \emph{acotxador}, respectively). We make this simplification because these are the main important floors in the \emph{pilars} and the situations that we are trying to describe in these examples, at least in a first approach, and it suffices to give some insight in the approach here considered. But, of course, this analysis should be performed for the whole \emph{pilar} in a future work. 


\begin{table}[h!]
\centering
 \begin{tabular}{||c |l |c | c c c||}
 \hline
 \emph{Casteller} & Joint angle & CJXdT & CdV (frame 1) & CdV (frame 2) & CdV (frame 3) \\ [0.5ex]
 \hline\hline
 \multirow{3}{*}{\emph{Quart} ($i=1$)} & feet ($\theta_{11}$) & 0 & 0 & 0 & 0\\ 
                        & knee ($\theta_{12}$) & 0 & 0 & 0 & 0\\ 
                        & hip ($\theta_{13}$) & $-11$ & $-12$ & $-16$ & $-32$ \\
 \hline \hline
 \multirow{3}{*}{\emph{Quint}  ($i=2$)} & feet ($\theta_{21}$) & $-5$ & $-15$ & $-21$ & $-40$ \\
                        & knee ($\theta_{22}$) & 32 & 20 & 23 & 35 \\
                        & hip ($\theta_{23}$) & $-12$ & $-13$ & $-16$ & $-20$ \\
 \hline \hline
 \multirow{3}{*}{\emph{Sis\`e}  ($i=3$)} & feet ($\theta_{31}$) & $-13$ & $-20$ & $-19$ & $-10$ \\
                         & knee ($\theta_{32}$) & 29 & 26 & 23 & 26 \\
                         & hip ($\theta_{33}$) & $-12$ & $-10$ & $-10$ & $-10$ \\
 \hline \hline
 \multirow{3}{*}{\emph{Acotxador} ($i=4$)} & feet ($\theta_{41}$) & $-5$ & $-$ & $-$ & $-$ \\
                            & knee ($\theta_{42}$) & 32 & $-$ & $-$ & $-$ \\
                            & hip ($\theta_{43}$) & $-12$ & $-$ & $-$ & $-$ \\
 [1ex]
 \hline
\end{tabular}
\\
\caption{Angle configuration (in degrees and with respect to the vertical) for the fixed configurations of the photos in the left of Figures \ref{JXT} and \ref{Vilaf}.}
\label{angles}
\end{table}

By looking at these specific frames, it seems quite clear that the one in Figure \ref{JXT} corresponds to a \emph{succesful} pillar (it did not collapse, indeed), while the three ones in Figure \ref{Vilaf} show the evolution of a pillar that ends up collapsing (with the last frame in Figure \ref{Vilaf} showing the precise moment previous to this collapse).
In both figures, the shape graphic (in the middle of these figures) is obtained by plotting the angle values given in Table \ref{angles}. The torque graphic (at the right of these figures) corresponds to the torque received by each joint, computed according formula \eqref{eq:torque} and using the angles values given in Table \ref{angles} and the biomechanical parameters given in \eqref{biomech}. 

The shape and torque graphics show that, as expected, the configuration corresponding to the three frames of Figure \ref{Vilaf} is worse than the one corresponding to Figure \ref{JXT}. And also that, among the three frames of Figure \ref{Vilaf}, the torque at certain joints and \emph{castellers} is increasing as the pillar is evolving and starting to collapse.
We can see in all cases that a critical joint seems to be the one between the \emph{quart} and \emph{quint} \emph{castellers}, which is actually the joint where the pillar of Figure \ref{Vilaf} ends up breaking (see the link to the video given at the caption of Figure \ref{Vilaf}).

With all these data we compute the first two quality indicators seen above (eccentricity and folding index, see \eqref{eq:eccentricity} and \eqref{eq:folding}, respectively). In future works it would be interesting to contact biomechanical experts in order to obtain the maximum torque allowed for an average feet, knee or hip joint, which is needed  to compute the suffering index. 

The results for the eccentricity and folding index can be seen in Table \ref{taula}. Note that the computations related with the three frames do not take into account the effect of the \emph{castellers} that are descending. (which, of course, is a very important disrupting factor for the stability of the structure and should be considered in a further work). This is also the reason for not having included the \emph{acotxador} rotation in Table \ref{angles}.

\begin{table}[h!]
\centering
 \begin{tabular}{||c |l |c | c c c||}
 \hline
 \emph{Casteller} & Joint angle & CJXdT & CdV (fm 1) & CdV (fm 2) & CdV (fm 3) \\ [0.5ex]
 \hline\hline
 \multirow{5}{*}{Mass center eccentricity} & \emph{Quart} ($i=1$) & $0.0310$ & $0.1240$ & $0.1718$ & $0.2820$ \\ 
                        & \emph{Quint} ($i=2$) & $-0.1365$ & $-0.0156$ & $-0.0096$ & $-0.0908$ \\ 
                        & \emph{Sis\`e}  ($i=3$)& $-0.0638$ & $-0.1300$ & $-0.1889$ & $-0.4075$ \\
                        & \emph{Acotxador} ($i=4$)& $-0.1099$ & $-$ & $-$  &$-$ \\ 
                        & \textbf{TOTAL} & $\mathbf{0.0310}$ & $\mathbf{0.1240}$ & $\mathbf{0.1718}$ & $\mathbf{0.2820}$\\
 \hline \hline
 \multirow{5}{*}{Folding index} & \emph{Quart} ($i=1$) & 0.0000 & 0.0000 & 0.0000 & 0.0000\\ 
                        & \emph{Quint} ($i=2$) & $0.2332$ & $0.3175$ & $0.4475$ & $0.8848$\\ 
                        & \emph{Sis\`e}  ($i=3$) & $0.1829$ & $0.5554$ & $0.7554$ & $1.1544$ \\
                        & \emph{Acotxador} ($i=4$) & $0.2106$ & $-$ & $-$ & $-$\\
                        & \textbf{TOTAL} & $\mathbf{0.6267}$ & $\mathbf{0.8729}$ & $\mathbf{1.2030}$ & $\mathbf{2.0391}$\\
 [1ex]
 \hline
\end{tabular}
\\
\caption{Summary of quality indicators for Figures \ref{JXT} and \ref{Vilaf}.}
\label{taula}
\end{table}

The values in Table \ref{taula} obtained in each case support the idea that these are indeed good indicators of the stability of a \textit{pilar} and correlate positively with the external visual perception and the internal sensing of the quality of a \emph{castell}. Indeed, it can be seen that all the indicators are lower in the case of the \emph{good} pillar of CJXdT (Figure \ref{JXT}) and higher as the pillar of CdV is getting worse (Figure \ref{Vilaf}), both in each flor or for the whole \emph{castell}. The last frame of this pillar (corresponding to the moment previous to the collapse) has the highest values of these quality indices. In addition, the eccentricity index corresponding to the \emph{quart} of the pillar of CdV is also the highest one, and this also correlates with the fact that the breaking of the pillar is produced at the joint between the \emph{quart} and \emph{quint}.

These computations have been done using averaged, simulated or computed values of the involved parameters and data, to use the pillars of Figures \ref{JXT} and \ref{Vilaf} as paradigmatic examples of the suggested approach, which seems to correlate with experience of \emph{castells}. But, of course, and as a further work, it would be important to use proper biomechanical and physical data to compute the torques and the quality indices. As some of them may not be known, this future work would imply the design of experiments in order to obtain them.

\section{Conclusions and future directions}\label{sec:future}
Understanding which are the mechanics and mechanisms that allow a \emph{castell} to be built and maintained is the first step for understanding the causes of its collapse and how and where do the \emph{castellers} fall down once the structure is broken, and the main goal of this first work on the subject. Among all, the \emph{pilar} is a type of \emph{castell} that includes most of the important challenges to deal with in a first mechanical model. That is why we have focused our study in the \emph{pilar} from three different points of view: static, dynamical and control model.

First, we have considered the dynamical problem thinking the \emph{pilar} as $N$-link inverted pendula (each of the links corresponding to one \emph{casteller}), when assuming passive (non reacting) \emph{castellers}. Design of experiments in order to calibrate the model parameters should be designed. Also, a better understanding of the underlying dynamics of this model would be the first step to understand the physical reasons of a \emph{pilar} collapse and, hence, to develop a first-failure model with impacts.

Second, assuming active response \emph{castellers} would mean to consider the control problem. Also, to make this problem more realistic, we have considered a 3-link \emph{casteller} model, where the movement of the joints of each \emph{casteller} can be used as a control of the system. As a starting point for this model, we have considered the case of a single \emph{casteller} with a passivity-based controller to achieve the desired position, where the interaction with the \emph{castellers} of other floors not considered. The 3-link \emph{casteller} has introduced the concept of internal joints, which represent the body articulations (in this case, knees and hip) and external ones (in this case, the joint between the feet of one \emph{casteller} and the \emph{casteller} below him). While the external ones can break (and this should be implemented in a future work), the internal ones cannot. Instead, they show an asymmetric behaviour as they cannot bend past a certain limit angle. In order to account for this, we have introduced a kinematic constraint into the model. We have considered a three conditions balance criterion for the 3-link \emph{casteller}, which is related with the controller: align the centre of mass horizontally with the feet, align the shoulders horizontally with the feet and bring the centre of mass to a given height, which should represent a ``comfortable'' position for the \emph{casteller}. We also introduced a delay-differential balance model which, even though we do not explore in this work, it could be included to account for the reaction time of a \emph{casteller}.

The 3-link \emph{casteller} model have arisen because, as we can observe in Figure \ref{fig:diagram_pilar}, \emph{castellers} adopt a bent-knee position rather than standing perfectly vertical. This fact posed the question of what are the benefits of this position. In order to analyse this position, we have considered a 2-link model with an asymmetric joint and we ran simulations for three different desired positions (standing, small bending and large bending). We have found that, starting each simulation at the corresponding equilibrium position with the same initial velocity, the more bent the position the faster it returned to equilibrium, even though it implied larger control torques at early times.

Again in this approach, the control parameters should be calibrated, first for an average configuration data (control to achieve a certain goal configuration) and, second, adapt them to dynamical data (control when the \emph{pilar} is moving). In the first part, we should consider things such as biomechanics (what human body is able to do in order to achieve a certain position, which will give kinematic constraints to the problem) or the quality of a \emph{casteller} (control parameters may be different depending on the ability of a \emph{casteller}). In the second part, it would be important in future works to develop aspects such as the delay in the response to an external stimulus or the fatigue that the \emph{castellers} are experiencing as the \emph{castell} goes on (including them as a delay problem and, maybe, considering that values of the control parameters could change with time).

Further work on this control problem approach would include improving the control strategy by, for instance, implementing joint-coordinated control strategies for each \emph{casteller} (control strategies taking into account all the joints of a \emph{casteller} and not thinking them as independent) to decide its own control action, as well as incorporating in them the interaction with the rest of the floors. These joint torque control strategies should account, of course, for realistic dynamical feedback features of the human neuro-muscular system, including fatigue effects, as said above. Also, it would be very interesting (and challenging) the study of the static and dynamic stability of the realistic controlled model. One may think that several equilibria are possible, but not all of them with the same stability nor with the same cost to be achieved or maintained.

Finally, we have approached the mechanics in the static problem for the whole \emph{pilar} considering again a 3-link \emph{casteller} model: given a fixed configuration, we have obtained the torques at each joint of the \emph{castellers} considering their own moments and those coming from the interaction with the other floors in the whole \emph{pilar}. This has allowed us to compute three quality indicators that seem to correlate with the feasibility of a \emph{castell}, and that complement the balance criterion used in the control problem above. The relation between feasible torques and angles should be investigated, as it may provide feasible zones for a successful \emph{pilar}. It can also be used in order to validate an angle configuration given by the solutions of the dynamical and/or control problems at a fixed time.

In the previous discussions we have provided some simplified examples to give some insight on our suggested approaches. But, in order to have feasible results, proper and accurate biomechanical data should be used, such as maximum torques or angles allowed by particular joints in human body or lengths, mass densities and mass centers of human joints. Hence, we should look for the type of data that is already known in the scientific literature and design possible studies in order to collect the unknown one, taking into account biomechanical differences between \emph{castellers} of different floors. Also, the design of a studies on how to compute and collect accurate angles of a configuration of a \emph{pilar} at a fixed time (possibly involving computer vision techniques) should be considered.

Finally, future modelling of a collapsing \emph{castell}, incorporating joint ruptures and impact phenomena, will ask for a more involved dynamical model and efficient numerical techniques. Also, in order to produce useful results, with respect to the end-problem addressed, a realistic dynamical computational tool should be produced, able to achieve a large number of numerical simulations, from which statistical results may be inferred concerning the most probable \emph{castellers} 
``landing'' locations, body positions and impact forces. In this work we have considered the \emph{pillar} as a case study for the \emph{castells} modelling, as it can be thought as a 1D-structure that has a 2D-movement in most of the situations. But we think the approaches and techniques described in this report in the case of a \emph{pilar} are the first step to the modelling of a \emph{castell} as a 3D complex structure. New challenging aspects such as, for instance, torsion within each floor should be then considered in future works.

\section{Acknowledgements}
This problem was suggested at the 158th European Study Groups with Industry that took place at the Centre de Recerca Matem\`atica (Universitat Aut\`onoma de Barcelona) from 27th to 31st of January 2020 by the Coordinadora de Colles Castelleres de Catalunya (CCCC). The authors would like to thank the CCCC, and specially its Scientific and Medical coordinator Dr. Dani Castillo, for this. Also, we would like to thank Dr. Jaume Rosset and Xavi Rius for their interesting and helpful discussions on the problem.

This work is partially supported by the following grants. F. Brosa Planella has been supported by The Faraday Institution (faraday.ac.uk; EP/S003053/1), grant number FIRG003. A. D\`{o}ria-Cerezo has been partially supported by the \emph{Generalitat de Catalunya} through the Project 2017 SGR 872. M. Pellicer is part of the Catalan Research group 2017 SGR 1392 and has been supported by the MINECO grant MTM2017-84214-C2-2-P (Spain).
\appendix
\section{Short Glossary}\label{sec:appendix}
\begin{figure}[htpb]
\begin{center}
\includegraphics[scale=0.2]{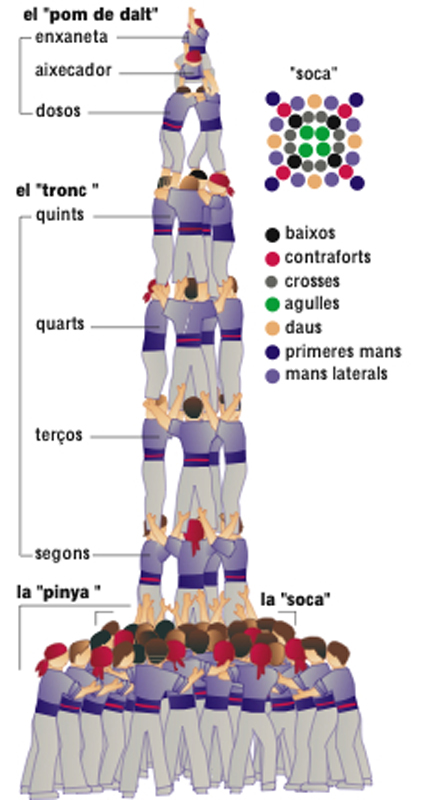}
\caption{Structure of the \emph{castell} known as \emph{quatre de 8 (4d8)}, indicating the parts of the \emph{castell} and positions of the \emph{castellers} (photo from Fototeca.cat).}\label{partscastell}
\end{center}
\end{figure}

In this section we introduce some of the terminology about \emph{castellers} that has been used in this report (note that an `s' at the end of a word denotes the plural form). See also Figure \ref{partscastell}.

Each \emph{castell} is named by the number of floors ($N$) and the number of \emph{castellers} in each floor ($M$) and is usually abbreviated as $MdN$. For example, a \emph{4d8} stands for a \emph{castell} of $8$ floors, with $4$ people in each floor. When we have only one person per floor ($M=1$) we name the \emph{castell} as \emph{pilar}. For instance, a \emph{pd6} stands for a \emph{castell} of $6$ floors, with $1$ person in each floor.

\vspace{0.1cm}
\noindent\textbf{Acotxador:} \emph{casteller} (usually a child) that occupies the second-to-top level of a \emph{castell}.

\noindent\textbf{Ball de valencians:} translated as Dance of the Valencians, they are believed to be the performances that originated the \emph{castells} in the XVIII century.

\noindent\textbf{Carregat/Descarregat:} it is said a \emph{castell} is \emph{carregat} once the \emph{enxaneta} reaches the top of it. The \emph{castell} is considered to be \emph{descarregat} once it is successfully disassembled.

\noindent\textbf{Castell:} human tower, of variable size and height, made of \emph{castellers} standing on each other's shoulders. Typical from traditional festivals in Catalonia.

\noindent\textbf{Casteller:} person that takes part in the building of \emph{castells}.

\noindent\textbf{Colla:} teams or groups that perform the \emph{castells}, each of them of a different town, village or neighbourhood. They can be distinguished by the colour of their shirt.

\noindent\textbf{Enxaneta:} \emph{casteller} (usually a child) at the top of a \emph{castell}. Once he or she has reached and crossed the top of the structure, the \emph{castell} is considered to be \emph{carregat}.

\noindent\textbf{Folre:} Second basis that needs to be build on the \emph{pinya} in higher \emph{castells} to keep them stable.

\noindent\textbf{Manilles:} Third basis that needs to be build on the \emph{manilles} in even higher \emph{castells} to keep them stable.

\noindent\textbf{Pilar:} (or pillar) \emph{castell} made of a single \emph{casteller} at each level.

\noindent\textbf{Pinya:} group of people that hold the basis of a \emph{castell}.

\noindent\textbf{Pom de dalt:} the top three levels of a castell (unless it is a \emph{pilar}), with the \emph{enxaneta} at the top one and the \emph{acotxador} at the second one.

\noindent\textbf{Quart:} \emph{casteller} that is in the fourth floor of a \emph{castell}.

\noindent\textbf{Quint:} \emph{casteller} that is in the fifth floor of a \emph{castell}.

\noindent\textbf{Sis\`e:} \emph{casteller} that is in the sixth floor of a \emph{castell}.

\noindent\textbf{Tronc:} vertical part of a \emph{castell}.


\end{document}